\documentclass[twocolumn]{aastex63}
\usepackage{float}
\usepackage[utf8]{inputenc}

\usepackage{graphicx}
\usepackage{hyperref}
\usepackage{amsmath}
\usepackage{color}
\usepackage{xcolor}
\usepackage{enumitem}
\usepackage{soul}
\usepackage{amssymb}
\usepackage{pifont}
\newcommand{\xmark}{\ding{55}}%

\hypersetup{colorlinks=true, citecolor=cyan, urlcolor=blue, linkcolor=blue}

\newcommand{\eg}{e.\,g.}
\newcommand{\ts}{\textsuperscript}
\newcommand{\prob}{\text{I\kern-0.15em P}}

\DeclareUnicodeCharacter{2212}{-}

\shorttitle{Fourier versus Bayesian Periodicity Detection for SMBHBs}
\shortauthors{Banaszak et al.}

\received{\today}
\submitjournal{\apj}

\begin{document}
\title{
Optimizing Optical Searches for Supermassive Black Hole Binaries in AGN Light Curves: Fourier versus Bayesian Periodicity Detection}

\correspondingauthor{Sebastian M. Banaszak}
\email{sebastian.m.banaszak@vanderbilt.edu}

\author[0009-0003-7203-1228]{Sebastian M. Banaszak}
\affiliation{Vanderbilt University, 2201 West End Ave, Nashville, TN 37235, USA}

\author[0000-0002-6020-9274]{Caitlin A. Witt}
\affiliation{Department of Physics, Wake Forest University, 1834 Wake Forest Road, Winston-Salem, NC 27109, USA}
\affiliation{Center for Interdisciplinary Exploration and Research in Astrophysics (CIERA), 1800 Sherman Ave., Evanston, IL 60201, USA}
\affiliation{Adler Planetarium, 1300 S. DuSable Lake Shore Dr., Chicago, IL 60605, USA}

\author[0000-0001-9515-478X]{Adam~A.~Miller}
\affiliation{Department of Physics and Astronomy, Northwestern University, 2145 Sheridan Road, Evanston, IL 60208, USA}
\affiliation{Center for Interdisciplinary Exploration and Research in Astrophysics (CIERA), 1800 Sherman Ave., Evanston, IL 60201, USA}
\affiliation{NSF-Simons AI Institute for the Sky (SkAI), 172 E. Chestnut St., Chicago, IL 60611, USA}

\begin{abstract}

Simulations predict that supermassive black hole binaries (SMBHBs) will exhibit periodic brightness variations that may exceed the stochastic variability intrinsic to active galactic nuclei (AGN). In this paper, we simulate SMBHBs with damped random walk (DRW) AGN variability and an added sinusoidal signal from the orbital motion, and test three methods—the Generalized Lomb Scargle Periodogram (GLSP), the nested Bayesian sampler (NBS), and the Weighted Wavelet Z-Transform (WWZ)—to determine which is best at recovering the periodicity. Our simulated light curves follow the properties of the Catalina Real-Time Transient Survey (CRTS), Legacy Survey of Space and Time (LSST), and Zwicky Transient Facility (ZTF) to best inform current and future SMBHB searches. We map a broad range of parameter space and identify which DRW-only light curves best mimic periodicity and pass each method’s model selection. The NBS performs best at detecting periodicity and filtering out DRW-only light curves. Combined candidate selection with both the NBS and GLSP significantly reduces false positive rates with marginal impact to true positive rates. With this joint model selection pipeline, we find the lowest false positive rates in ZTF-like simulations and the highest detection rates in LSST-like simulations. Using a modified computation of the False Alarm Probability (FAP) with GLSP, we efficiently triage LSST AGN light curves ($\sim$$10^7$ light curves in $\sim$10--30 hours) and achieve true- and false- positive rates of $\sim$40\% and $\sim$0.5\%, respectively.

\end{abstract}


\section{Introduction} \label{sec:intro}
Several theoretical models predict the existence of supermassive black hole binaries (SMBHBs) following hierarchical galaxy mergers, which are common in cosmological $\Lambda$CDM simulations \citep{magorrian1998_smbh_ubiq, haehnelt2002_hierarchical_smbh, springel2005_gal_sims}. According to these models, SMBHBs produce gravitational wave (GW) \textit{and} electromagnetic (EM) signals, making them an outstanding candidate for EM+GW multi-messenger astrophysics (MMA) research. Given the conditions of their formation, the detection of SMBHBs presents novel opportunities to study a wide range of phenomena, including but not limited to: SMBH-galaxy coevolution, hierarchical galaxy formation, general relativity and accretion physics in turbulent environments.

Pulsar Timing Arrays (PTAs) around the world have found evidence for a (Nanohertz) Gravitational Wave Background which is likely composed of inspiraling SMBHBs \citep{
agazie2023_15yrCW, epta_dr2, reardon2023_gwb_noise, xu2023_gwb_cpta}. Continuous GWs from individual SMBHBs could be detected above the stochastic GW background within the decade \citep{rosado2015_cws, mingarelli2017_localgwb_cw, kelley2018_cw, taylor2020_cws_gwskies, becsy2022_building_gwb}, and such detections would enable us to identify individual SMBHB systems through GW observations \citep{kato2023_localize_gw, petrov2024_hostgal_followup}. Pairing PTA methods with EM detection methods provides us a more complete and accurate picture of SMBHB systems \citep{3c66b, xin2021_mma_crts, liu2021_mma, liu2023_mma, charisi2025_mma}. In this paper, we investigate an EM detection method and consider our results in the context of a multi-messenger search for SMBHBs.

Closely-separated SMBHBs may be detected as luminous Active Galactic Nuclei (AGN) with periodically variable brightness \citep{haiman2009_smbhb, tanaka2012_smbhb_as_agn, bogdanovic2022_em_smbhb}. Relativistic effects and interactions between three bright accretion disks (a circumbinary (CBD) disk and two ``mini-disks" around each SMBH) induce periodic variations in the system's brightness \citep{shi2012_3dmhd_cbdisks, farris2014_mbhbsims, dorazio2015_rdb_smbhb, Tang+2018, saeedzadeh2024_dual_agn}. When SMBHB brightness is modulated by a relativistic Doppler boost (RDB), detection of sinusoidal periodicity in an AGN light curve thus serves as evidence for binary motion in an SMBHB. Stochastic brightness variations intrinsic to AGN make it difficult to extract this periodicity, especially in sparsely sampled, short-baseline light curves. This emphasizes the need to carefully calibrate the False Positive Rate (FPR) in searches for quasar periodicity \citep{vaughan2016_fpr, barth2018_fpr, liu2019_panstarrs_candidate, witt2022_agnvar_bayesian}.

Other EM detection methods involving observations across the entire electromagnetic spectrum have been investigated in detail \citep{bogdanovic2009_smbhb_spec, boroson_lauer_2010_smbhb_spec, tsalmantza_2011_smbhb_spec, ju2013_smbhb_spec, decarli_2014_smbhb_spec, liu_shen_2014_smbhb_spec, runnoe_2017_smbhb_spec}. Continuum reverberation mapping and searches for periodically shifting emission lines in AGN spectra require intensive treatment of degeneracies with spectrum-wide shifts and emission line broadening. Furthermore, spectroscopic follow-up will be especially demanding in this age of photometric-dominated surveys. So, while these methods will likely provide invaluable follow-up at strong SMBHB candidates, time series observations provide the most promising avenue to large-scale SMBHB detection. In fact, SMBHB candidates have been identified in several optical surveys, including the Catalina Real-Time Transient Survey (CRTS) \citep{Graham2015_crts_candidate, liu2015_crts_candidate}, the Palomar Transient Factory \citep{Charisi2016_agnvar}, the Panoramic Survey Telescope and Rapid Response System \citep{liu2019_panstarrs_candidate}, a combination of the Sloan Digital Sky Survey and Dark Energy Survey \citep{chen2020_des_sdss_candidate}, and the Zwicky Transient Facility (ZTF) \citep{chen2024_agnvar_ztf}. Most excitingly, the Legacy Survey of Space and Time (LSST) at the Vera C.~Rubin Observatory will observe between $20$ and $100$ million quasars in the $ugrizy$ filters over a $10$\;yr baseline \citep{ivezic2019_lsst_stuff, kelley2019b_smbhb_periodic, xin2021_pop_estimates, charisi2022_agnvar}.\footnote{\url{https://pstn-052.lsst.io/}} Based on simulations, several hundred binary candidates will be observable in LSST \citep{ivezic2017_lsst_agn, kelley2019b_smbhb_periodic, xin2021_pop_estimates, kelley2021_gravlensing_smbhb}. Given this vast repository of existing and upcoming photometric AGN observations, consistent development of PTA detection methods, and opportunities for new science, a robust quasar periodicity detection pipeline is especially opportune.

In this paper, we test three methods' ability to detect periodic variability among stochastic noise in simulated AGN light curves. We use Generalized Lomb Scargle Periodograms (GLSP) implemented in the \texttt{PyAstronomy} package \citep{pyAstronomy}, the nested Bayesian sampler (NBS) from the \texttt{Ultranest} package \citep{ultranest}, and the Weighted Wavelet Z-transform (WWZ) algorithm implemented in the \texttt{wwz} package \citep{wwz_sebastian}.\footnote{\url{https://pyastronomy.readthedocs.io/en/latest/pyTimingDoc/pyPeriodDoc/gls.html}}\footnote{\url{https://johannesbuchner.github.io/UltraNest/}}\footnote{\url{https://github.com/skiehl/wwz/tree/main}} We simulate a population of idealized AGN light curves using a ``damped random walk" (DRW) model to account for stochastic variations, and a sinusoid to account for periodic variations due to periodic RDB in an SMBHB orbit. Each light curve is simulated as it would appear in CRTS, ZTF, and LSST light curve data. We construct a pipeline which employs Bayesian and Fourier model-selection and parameter-estimation to identify and characterize periodic signals in our sample. Finally, we evaluate the ability of each methods to detect periodicity in AGN light curves and filter out false positives, noting best uses for each method in the context of a data-heavy search for SMBHBs.




This paper is laid out as follows. In section 2, we describe the methodology for simulating AGN light curves and the details of parameter estimation and model selection for each method. In section 3, we evaluate the fidelity of each method and discern their best uses. In section 4, we discuss future improvements to our methods and contemporary work in EM+GW MMA. In section 5, we draw notable findings from our results and interpret them in the context of searches for SMBHBs in real data. This work represents an early development in the electromagnetic side of a joint EM+GW SMBHB detection pipeline.
\section{Methods} \label{sec:methods}
To test parameter estimation and model selection with the NBS, GLSP and WWZ, we simulate CRTS, LSST, and ZTF light curves of SMBHs and SMBHBs using realistic and predicted properties of AGN and SMBHBs. With this population, we constrain the genuine true positive rates (TPR($P$)) and false positive rates (FPR) expected in a systematic search for periodicity in AGN light curves (See \autoref{subsec:binary_class} for rate definitions). We also investigate the possibility of a rapid triage for the large volume of light curves expected from LSST (\autoref{subsec:rapid_fap}).

\subsection{Simulating Light Curves}\label{subsec:sims}
To produce realistic light curves, we:

\begin{enumerate}
    \item Simulate AGN variability using the DRW model with realistic timescale and variability parameters.
    \item Inject sinusoid signals into $\sim$half of the light curves. These represent the SMBHB light curves in our population.
    \item Sample the simulated time series to match the observational cadence, mean magnitude and mean photometric uncertainty of CRTS, LSST, and ZTF light curves.
\end{enumerate}

\subsubsection{Survey Properties}\label{subsubsec:surveys}
We extract 10,000 AGN light curves spread across the sky from CRTS\footnote{CRTS; \url{http://nesssi.cacr.caltech.edu/catalina/AllAGN.arch.html}} \citep{drake2009_crts} and ZTF\footnote{ZTF; \url{https://www.ztf.caltech.edu/ztf-public-releases.html}} \citep{bellm2019_ztf} online databases. This includes observation dates (MJD; Modified Julian Date), magnitudes, and photometric uncertainties.

For LSST cadencing and observation baselines, we use the \texttt{rubin\textunderscore sim} package from the \texttt{lsst} framework to simulate full-length LSST light curves across the survey's coverage area. This framework incorporates a recent baseline survey (v$3.4$), according to which LSST will employ a ``Wide-Fast-Deep" (WFD) observing strategy over a $\sim$18,000 deg$^{2}$ region.\footnote{LSST; \url{https://survey-strategy.lsst.io/baseline/index.html}. The most recent baseline survey, v$4.1$, introduces changes which will not substantially affect our results.} We refer to Table $2$ and equations ($4$) and ($5$) of \citet{ivezic2019_lsst_stuff} to generate mean magnitudes and photometric uncertainties for each simulated light curve---a single mean photometric uncertainty was applied to all observations in a given light curve.

We bin-average MJD for observations taken in the same night to avoid sampling light curves at high cadences ($\Delta t \leq 1$ day), where the DRW fails to model AGN stochastic variability \citep{zu2013_drw_short_timescale}. We require at least $100$ bin-averaged observations for ZTF light curves, $30$ for CRTS light curves and $45$ for LSST light curves. CRTS, LSST, and ZTF light curve properties are summarized in \autoref{tab:survey_info}. We select a random subset of $\sim$3500 CRTS, LSST, and ZTF light curves and use the observation dates, mean magnitude and photometric uncertainties as templates for simulations.

\begin{table}[ht!]
    \centering
    \caption{Average Parameters Used to Simulate Data Sets}
    \begin{tabular}{ccccc}
         \hline
         Survey & Baseline & Cadence & Mean Phot. & Observation \\
         & (years) & (days) & Unc. & Count \\
         \hline
         \hline
         CRTS & $8-11$ & $10-30$ & $0.14$ & $74$ \\
         LSST & $9.5-10$ & $5-10$ & $0.07$ & $134$ \\
         ZTF & $5.2$ & $3-5$ & $0.08$ & $282$ \\
         \hline
    \end{tabular}

    \label{tab:survey_info}
\end{table}



\begin{figure*}[ht!]
    \centering
    \includegraphics[width =1\linewidth]{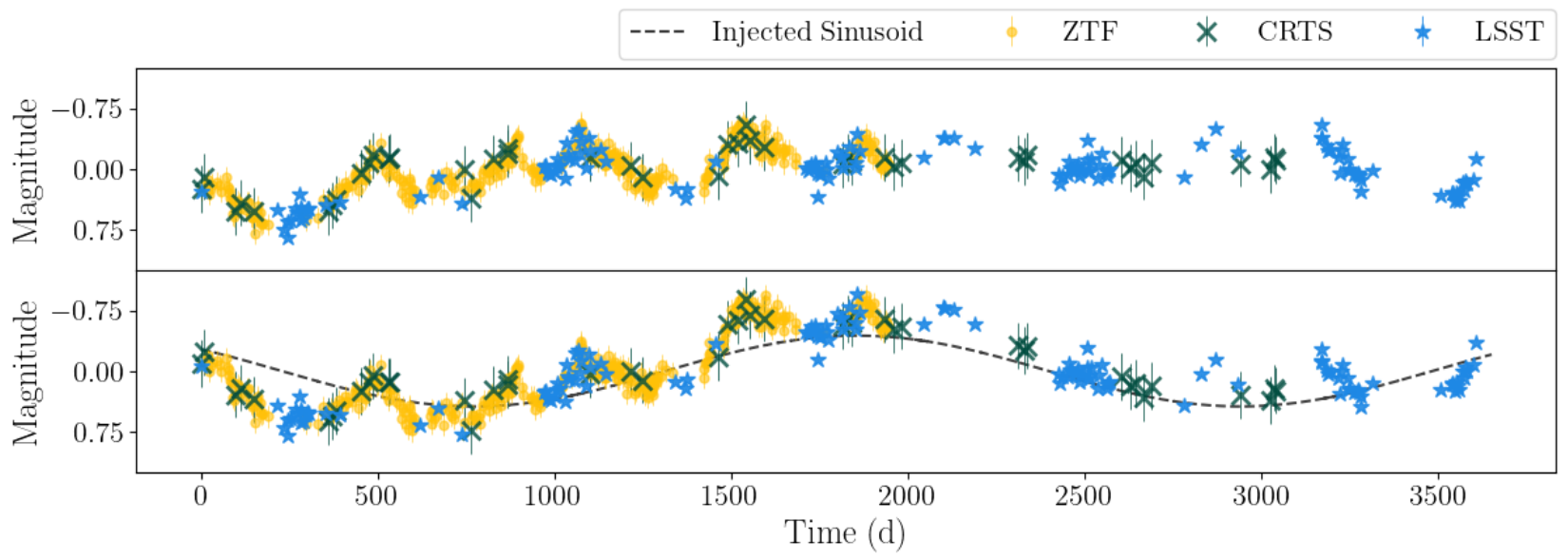}
    \caption{Simulated light curves containing a DRW process (top panel) and the same DRW process plus a sinusoid (bottom panel, shown in dashed gray). Simulated observations are shown for CRTS (green X's), LSST (blue stars), and ZTF (closed yellow circles). Photometric uncertainty is for a sample 20-magnitude light curve. Note, light curves in CRTS, LSST and ZTF do not overlap as shown\--x-axis shows time since first observation.}
    \label{fig:lcs}
\end{figure*}

\subsubsection{DRW and Periodic Variability}\label{subsubsec:drw}

With observation templates prepared, we simulate single and binary AGN light curves following the steps from \citet{Charisi2016_agnvar}. The power spectral density (PSD) function of a DRW process is

\begin{equation}
    P(f) = \frac{4 \sigma^{2} \tau}{1 + (2 \pi \tau f)^{2}},
\end{equation}
where $\sigma^{2}$ is the variance of the light curve data points, $\tau$ is a characteristic DRW timescale, and $f$ is the Fourier space frequency. We select $\tau$ and $\sigma$ using the observed distributions in \citet{macleod2010_agnvar_drw} and generate evenly sampled light curves ($\Delta t$ = 1 day). We add white noise with a Gaussian distribution to account for measurement uncertainty.

The signal we use to simulate periodic variability has the form

\begin{equation}
    \mathbf{s}(t) = A \sin\left(\frac{2 \pi}{P}(t_{0} - t)\right),
\end{equation}
where $A$ is the amplitude in magnitudes, $P$ is the period in days, and $t_{0}$ is the reference time in days. The range of sinusoidal forms we use in our simulations are not explicitly connected to expectations of SMBHB system parameters. They represent a variety of signals which correspond to different types of periodic variability which may occur in an SMBHB. Our sinusoidal signals most directly correspond to Relativistic Doppler Boosting (RDB) variability for an uneven-mass, circular-orbit SMBHB. Considerations of alternative periodic signal forms will be discussed in \autoref{sec:discussion}.

The periods of the injected sinusoids range from 30 days to 10 yr, capped at the most common LSST dataset baseline. This wide range of periods largely overlaps with GW frequencies detectable by PTAs. With this range, we don't account for SMBHBs possibly detectable by the Laser Interferometer Space Antenna (LISA; \citealt{xin2021_pop_estimates}), which are expected to have periods of only a few days. As we demonstrate below, short period binaries are easier to detect so we are optimistic that LISA SMBHBs will be recoverable provided short-timescale AGN fluctuations can be modeled.

\begin{table*}
\centering

\caption{Simulation Ranges, Bayesian Prior Shapes, and Bootstrap Simulation Distributions for Our Five Parameters}
    
    \begin{tabular}{cccc}
        \hline
        {Parameter} & Simulation Distribution & Prior & Bootstrap/Database Simulations\\
        \hline
        \hline
         $\mathrm{log}_{10}\sigma$ & Uniform[$-1.6$, $-0.25$]  & Uniform[$-1.6$, $-0.25$] & Uniform[$-1.6$, $-0.25$]\\
         $\mathrm{log}_{10}\tau $ & SkewNorm($3.0$, $0.5$, $-1.4$) & Uniform[$0.56$, $4.73$] & Uniform[$0.56$, $4.73$]\\
         $P$ & Uniform[$30$, $3652.5$] & Uniform[$30$, $3652.5$] & ... \\
         $A$ & Uniform[$0.05$, $0.5$] & Uniform[$0$, $0.5$] & ... \\
         $t_0$ & Uniform[$0$, $P$] & Uniform[$\mathrm{MJD}_{0}$, $\mathrm{MJD}_{0}+3652.5$] & ... \\
         \hline
    \end{tabular}
    \vspace{.25cm}
    
    \textbf{Note.} Simulation ranges, prior shapes, and bootstrap simulation distributions (\autoref{subsec:GLSP}) for each of our five parameters.
    \label{tab:priors}
\end{table*}
As the final step in creating simulated light curves we downsample each simulated time series to match the observational cadence of the relevant survey as described in \autoref{subsubsec:surveys}. Single and binary AGN light curves with example CRTS, LSST, and ZTF templates are displayed in \autoref{fig:lcs}.


\vspace{.25cm}
\subsection{Bayesian Likelihood and Sampling Methods}

The NBS efficiently explores likelihood over the entire parameter space and avoids effects induced by unevenly sampled time series. We use the nested-sampling Monte Carlo algorithm, MLFriends \citep{mlfriends1, mlfriends2}, using the \texttt{Ultranest} package \citep{ultranest} to sample the likelihood.\footnote{\url{https://johannesbuchner.github.io/UltraNest/}} In particular, for each simulated light curve, we sample two likelihoods: 1) A DRW likelihood (\autoref{eq:drwlike}), and 2) A combined DRW+Sine likelihood (\autoref{eq:drwsinelike}).

The DRW likelihood function marginalized over the mean of the light curve is given by

\begin{equation}\label{eq:drwlike}
\begin{split}
    P(\mathbf{y} \mid \mathbf{p}) \propto  |C|^{-1 / 2}  & \left|L^{T} C^{-1} L\right|^{-1 / 2} \\ & \times \exp \left(-\frac{\mathbf{y}^{T} C_{\perp}^{-1} \mathbf{y}}{2}\right),
\end{split}
\end{equation}
with $\mathbf{p}$ the set of model parameters, $\mathbf{y}$ the vector of observed magnitudes, $L$ a vector of ones with a length equal to the number of data points, and $C$ as the noise covariance matrix. With $\mathbf{s}$ as the input signal vector sampled at simulated time stamps, the likelihood for the DRW+sine model is

\begin{equation}\label{eq:drwsinelike}
\begin{split}
    P(\mathbf{y} \mid \mathbf{p}) \propto  &|C|^{-1 / 2}   \left|L^{T} C^{-1} L\right|^{-1 / 2} \\ & \times \exp \left(-\frac{(\mathbf{y}-\mathbf{s})^{T} C_{\perp}^{-1} (\mathbf{y}-\mathbf{s})}{2}\right).
\end{split}
\end{equation}

As shown in \autoref{tab:priors}, we use uniform and log-uniform priors on the model parameters. More informative priors could be imposed for the DRW parameters, in \citet{macleod2010_agnvar_drw} it is shown that $\sigma$ and $\tau$ are jointly correlated with the properties of the AGN (\eg, black hole mass, luminosity, etc.). We do not vary luminosity-related parameters in our simulations, so there is not a natural way to incorporate these correlations into our analysis. Note, the upper prior limit for the $P$ parameter occasionally exceeds the observation window. Since this doesn't systematically bias posteriors for light curves with $P_{\mathrm{in}}<T_\mathrm{obs}$, we permit this wide prior.

We take the median of the Bayesian posterior distributions for the model parameters to be the best-fit parameter values. Associated lower-bound and upper-bound uncertainties are computed as the 5\ts{th} and 95\ts{th} percentiles of these distributions. To determine which of the two models is preferred, we use the Bayes information criterion (BIC),

\begin{equation}
    \mathrm{BIC}=k \ln (n)-2 \ln (\widehat{L}),
\end{equation}
where $k$ is the number of free parameters, $n$ is the number of observations in the light curve, and $\widehat{L}$ is the maximum likelihood value \citep{liddle}. As shown in \citet{witt2022_agnvar_bayesian}, the BIC allows us to simply compare the models and avoids overfitting the data by accounting for the number of parameters in the model. In model comparisons, the model with the smallest BIC is usually preferred. We use this feature to define,

\begin{equation}
    \Delta \mathrm{BIC} = \mathrm{BIC_{DRW}} -  \mathrm{BIC_{DRW+Sine}}.
\end{equation}

A lower value of $\Delta \mathrm{BIC}$ thus indicates more support for the DRW+Sine model. In general, evidence for the DRW+Sine model can be considered positive for $−2 > \Delta \mathrm{BIC} > −6$, and strong for $\Delta \mathrm{BIC} < -6$ \citep{kass}.


\vspace{.25cm}
\subsection{Generalized Lomb-Scargle Periodogram Parameter Estimation and Model Selection} \label{subsec:GLSP}

The Lomb-Scargle Periodogram is an algorithm used for detecting and characterizing periodic signals in unevenly sampled time series \citet{lomb1976,scargle1982_gls}. The Generalized Lomb-Scargle Periodogram (GLSP) is an adaptation of the Lomb-Scargle Periodogram which includes measurement uncertainties and a non-zero mean. This generalization also provides a more accurate period prediction and a better determination of the spectral intensity than the LSP \citep{zechmeister2009_gls_stuff}. It is mathematically equivalent to fitting a sinusoidal function of form:

\begin{equation}
    s(t) = A \sin\left(\frac{2\pi t}{P} + \phi\right) + d,
\end{equation}
where $d$, $A$, $P$, and $\phi$ correspond to the mean, amplitude, period, and phase of a periodic signal, respectively. We use the GLSP algorithm implemented in the \texttt{PyAstronomy} package.

For each simulated light curve, we compute a periodogram at linear 1-day increments between periods $2\Delta t_\mathrm{revisit}$ and $T_{\mathrm{obs}}$, where $T_{\mathrm{obs}}$ is the observation window (days) and $\Delta t_\mathrm{revisit}$ is the median sampling cadence (days). Note, the GLSP can constrain some periodic signals with $P_{\mathrm{in}} > T_{\mathrm{obs}}$. However, since extending the period grid beyond $T_{\mathrm{obs}}$ systematically influences periodograms calculated for light curves with $P_{\mathrm{in}} < T_{\mathrm{obs}}$, we impose the strict upper limit of $T_{\mathrm{obs}}$.

We take the period corresponding to the largest periodogram peak to be the best-fit period and compute the best-fit amplitude, phase, and offset at this fixed period. For a full mathematical explanation of parameter estimation in GLSP, see \citet{zechmeister2009_gls_stuff}. We use the candidate periodogram---specifically, the largest periodogram peak height, $\mathcal{P}_{max}$---to perform model selection using the False Alarm Probability (FAP). The FAP measures the probability that a data set with no periodic signal would produce a periodogram peak of equal or greater magnitude, and amounts to estimating the significance of a periodogram peak. Here, we detail 3 variations of the FAP calculation, labeled $\mathrm{FAP_{Gauss}}$, $\mathrm{FAP_{local}}$, and $\mathrm{FAP_{global}}$.

\citet{scargle1982_gls} demonstrates periodogram values for a dataset of uncorrelated Gaussian noise follow independent, identical $\chi^{2}$ distributions with 2 degrees of freedom. $\mathrm{FAP_{Gauss}}$ is thus defined as,

\begin{equation}\label{eq:fap_gauss}
    \text{$\mathrm{FAP_{Gauss}}$} = 1 - \left(1 - e^{-\mathcal{P}_{max}}\right)^{M},
\end{equation}
the probability at least one of $M$ independent \textit{white-noise}-only periodogram values exceed $\mathcal{P}_{max}$. The \texttt{PyAstronomy} package approximates $M \approx f_{max}T_{\mathrm{obs}}$ to automatically compute $\mathrm{FAP_{Gauss}}$.


Assuming stochastic DRW variability dominates our time series data, periodogram values will follow correlated, non-identical distributions which are sensitive to survey window convolutions. Accommodating for these effects using a ``bootstrap" procedure allows us to more accurately discern detection significance. To implement the bootstrap procedure and calculate the FAP with respect to a DRW red noise background for each light curve, we:

\begin{enumerate}
    \item Simulate 20,000 DRW light curves, drawing from log-uniform $\sigma$ and $\tau$ distributions (\autoref{tab:priors}).
    \item Downsample each DRW simulation to match the candidate light curve's observation dates, mean magnitude, and photometric uncertainty.
    \item Compute the periodogram of each DRW simulation ($\mathcal{P}_{sim}\left(f\right)$), and record the power values at each frequency/period bin.
    \item Use the distributions of power values to computationally approximate FAP.
\end{enumerate}

Specifically, we define

\begin{equation}\label{eq:fap_global}
    \text{$\mathrm{FAP_{global}}$} = \frac{\mathcal{N}\left(max\left(\mathcal{P}_{sim}\left(f\right)\right) > \mathcal{P}_{max}\right)}{N_{sim}},
\end{equation}
as the probability a DRW light curve produces a more significant periodic signal detection than the candidate light curve. Then, we define

\begin{equation}\label{eq:fap_local}
    \text{$\mathrm{FAP_{local}}$} = \frac{\mathcal{N}\left(\mathcal{P}_{sim}\left(f_{max}\right) > \mathcal{P}_{max}\right)}{N_{sim}},
\end{equation}
to estimate the probability a DRW light curve produces a more significant periodic signal detection than the candidate light curve \textit{at the frequency of the detected signal}. Note, this means the $\mathrm{FAP_{local}}$ will be a poor estimate of detection significance if evaluated for the incorrect periodogram peak. On the other hand, it more accurately accounts for the uneven periodogram background produced by red-noise time series. We compare the two in \autoref{sec:results}.

While $\mathrm{FAP_{global}}$ and $\mathrm{FAP_{local}}$ are best suited for the problem at hand, we still calculate $\mathrm{FAP_{Gauss}}$ due to its tremendous computational efficiency. For the former two classifiers, $\left(1-\mathrm{FAP}\right) \geq 99.73\%$--a $3\sigma$-level detection--is typically considered as evidence for periodicity. For $\mathrm{FAP_{Gauss}}$, we manually select thresholds which yield comparably precise samples. 


\vspace{.25cm}
\subsection{Weighted Wavelet Z-transform Methodology and Parameter Estimation} \label{subsubec:WWZ}

The Weighted Wavelet Z-transform (WWZ) and Weighted Wavelet Amplitude (WWA) are improvements of the wavelet transform method, best suited for determining the period and amplitude of unevenly sampled time series \citep{foster1996_wwz}. Unlike the GLSP, the WWZ utilizes sine waves localized in both frequency and time, enabling it detect evolving periodic signals. While we expect unchanging periods in SMBHBs detectable with PTAs, this feature has been exploited to probe the temporary nature of quasi-periodic oscillations common in red noise datasets \citep{bhatta2016_candidate_glswwz, benkhali2020_candidate_glswwz, das2023_candidate_glswwz, gong2023_candidate_glswwz, zhang2024_candidate_gamma}.


In our implementation of WWZ, we adopt the abbreviated Morlet kernel which has the following functional form \citep{grossman_morlet1984_wwz}
\begin{equation}\label{eq:morlet}
    f\left[\omega\left(t-t_{s}\right)\right] = \exp\left[i\omega\left(t-t_{s}\right) - c\omega^{2}\left(t-t_{s}\right)^{2}\right],
\end{equation}
where $c$ is the decay constant. The WWZ map is then given as,
\begin{equation}\label{eq:wwzmap}
    W\left[\omega, t_{s}; x(t)\right] = \omega^{1/2}
    \displaystyle \int x(t) f^{*}\left[\omega\left(t-t_{s}\right)\right]\text{dt},
\end{equation}
where $x(t)$ is the light curve, $f^{*}$ is the complex conjugate of the Morlet kernel $f$, $\omega$ is the frequency, and $t_{s}$ is the time shift parameter.

For every simulated light curve, we compute the WWZ between $6\Delta t_\mathrm{revisit}$ and $T_{\mathrm{obs}}$ at 1 day increments, giving us $N_{\mathrm{WWZ}} = T_{\mathrm{obs}} - 6\Delta t_\mathrm{revisit}$ samples. The increased lower limit prevents WWZ from accumulating power at high frequencies for particularly sparse datasets \citep{foster1996_wwz}. For the time dimension, we compute the WWZ at 10 evenly-spaced times ($N_{t} = 10$) between the start and end observation date, which suits our relatively sparse datasets.

We take the period corresponding to the largest power spectrum peak to be the best-fit period. We compute the amplitude, phase and offset at the fixed best-fit period according to the same methodology as used for GLSP \citep{zechmeister2009_gls_stuff}.\footnote{Often, the WWA power value at the WWZ best-fit period and time is chosen as the best-fit amplitude. This value works well, but has a tendency to overestimate the amplitude and performs slightly worse than the alternative methodology we implement here.} We calculate the FAP to perform model selection for the WWZ. $\mathrm{FAP_{Gauss}}$ is not a valid metric for power spectra and is not calculated for WWZ. Instead, we use $N_{\mathrm{sim}} = 20,000$ bootstrap simulations to calculate $\mathrm{FAP_{global}}$ and slightly modified versions of \autoref{eq:fap_local}. With $\mathcal{P}_{sim}$ as the WWZ power spectrum of a bootstrap simulation, and candidate power spectrum peak $\mathcal{P}_{max}$ occurring at $\left(f_{max}, t_{max}\right)$, we define


\begin{equation}\label{eq:fap_local_local}
    \text{$\mathrm{FAP_{local\text{-}local}}$} =
    \frac{\mathcal{N}\left(\mathcal{P}_{sim}\left(f_{max}, t_{max}\right) > \mathcal{P}_{max}\right)}{N_{sim}}.
\end{equation}
Then, with $j \in \{1, ..., N_{t}\}$, we define

\begin{multline}\label{eq:fap_local_avg}
    \text{$\mathrm{FAP_{local\text{-}avg}}$} =\\ 
    \frac{1}{N_{t}} \displaystyle \sum_{j=1}^{N_{t}}\left(\frac{\mathcal{N}\left(\mathcal{P}_{sim}\left(f_{max}, t_{j}\right) > \mathcal{P}\left(f_{max}, t_{j}\right)\right)}{N_{sim}}\right).
\end{multline}
Also, with the average WWZ power spectrum, $\mathcal{P}_{avg}$ as

\begin{equation}\label{eq:avg_ps}
    \mathcal{P}_{avg}\left(f\right) = \frac{1}{N_{t}}\displaystyle \sum_{j=1}^{N_{t}}\mathcal{P}\left(f, t_{j}\right),
\end{equation}
we define,

\begin{multline}\label{eq:fap_avg_local}
    \text{$\mathrm{FAP_{avg\text{-}local}}$} =\\ 
    \frac{\mathcal{N}\left(\mathcal{P}_{sim}\left(f_{max}\right) > \mathcal{P}_{avg}\left(f_{max}\right)\right)}{N_{sim}}.
\end{multline}

The latter, $\mathrm{FAP_{avg\text{-}local}}$ is simply $\mathrm{FAP_{local}}$ calculated using the average WWZ power spectrum. $\mathrm{FAP_{local\text{-}avg}}$, deceivingly similar, is equivalent to taking the average of $\mathrm{FAP_{local\text{-}local}}$ computed at every time bin in the $f_{max}$ column. In the same way the GLSP $\mathrm{FAP_{local}}$ depends on the selection of the correct periodogram peak, $\mathrm{FAP_{local\text{-}local}}$, $\mathrm{FAP_{local\text{-}avg}}$ and $\mathrm{FAP_{avg\text{-}local}}$ depend on the selection of the correct power spectrum peak. If the peak is poorly localized in frequency \textit{or} in time, $\mathrm{FAP_{local\text{-}local}}$ will be a bad estimation of detection significance. $\mathrm{FAP_{local\text{-}avg}}$ and $\mathrm{FAP_{avg\text{-}local}}$ avoid issues of peak localization in time, because measures of significance are averaged along the time axis. For these FAP variations, $\left(1-\mathrm{FAP}\right) \geq 99.73\%$, a $3\sigma$-level detection, is typically considered as evidence for periodicity.

\subsection{LSST Survey Window and Rapid FAP Calculation}\label{subsec:rapid_fap}

\begin{figure}
    \centering
    \includegraphics[width=1\linewidth]{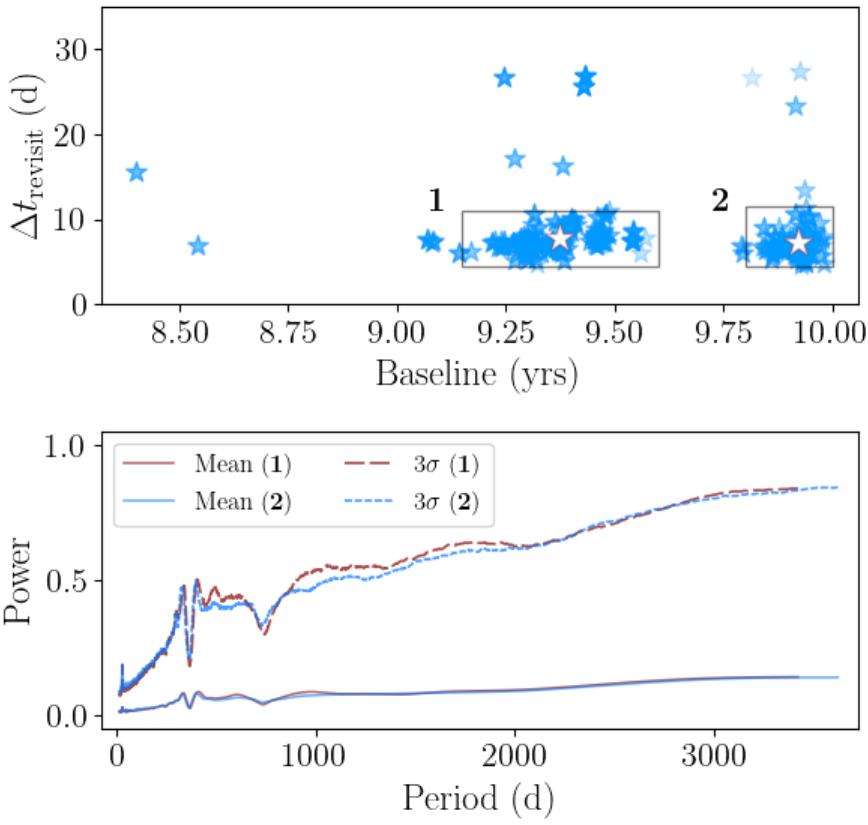}
    \caption{Median sampling cadence ($\Delta t_{\mathrm{revisit}}$) vs. baseline for LSST light curves (top panel), showing LSST survey windows will be systematically similar. Mean (solid) and $99.73$\ts{th} percentile (dashed) red noise periodogram values for ``representative" LSST survey windows (bottom panel). Representative survey windows (marked by red/white stars) are drawn from groups ``1" and ``2" labeled in top panel.}
    \label{fig:db_pvs}
\end{figure}

We expect a large portion of LSST light curves will have incredibly similar survey windows. In this case, the bootstrap simulation process---primarily used to account for effects caused by differences in observation cadence and duration---is slightly redundant. We evaluate a ``database procedure" in which, instead of rebuilding a database of periodogram value distributions for each candidate light curve, we build databases for just two ``representative" light curves and use them to compute the FAP for all candidate light curves. To build the databases, we follow the procedure detailed in \autoref{subsec:GLSP}, but instead of using the candidate light curve for downsampling, we use one of our two representative light curves. We use the mean LSST photometric uncertainty $\sigma_{err}=0.07$ for each DRW simulation. The mean and 99.73$\ts{th}$ percentile of each representative database are shown in \autoref{fig:db_pvs}. To estimate detection significance for a candidate light curve, we reference one of these two databases (depending on which group the candidate falls in) and compute $\mathrm{FAP}_{D_\mathrm{local}}$ and $\mathrm{FAP}_{D_\mathrm{global}}$ using \autoref{eq:fap_local} and \autoref{eq:fap_global}. The ``D" subscript denotes the classifier was computed with the database methodology.

In \autoref{sec:appendix}, we validate the database procedure against FAPs found with the bootstrap procedure and further investigate the causes for error. In \autoref{sec:results}, we discuss promising results which indicate $\mathrm{FAP}_{D_\mathrm{local}}$ could be implemented as a triage classifier.

\subsection{Binary Classification}\label{subsec:binary_class}


We use a modified binary classification (BC) model to evaluate the detection capacity of each method. Typically, a BC model requires that a candidate satisfies a single model selection criterion to be considered a ``true positive" detection. We further impose a parameter estimation criterion for a candidate to qualify as a ``genuine true positive". Note, we use this metric strictly to compare each method's detection capacity, and cannot apply it to a real SMBHB search since we do not know the intrinsic parameters. With $\Delta P = \left|P_{\mathrm{out}} - P_{\mathrm{in}}\right| / P_{\mathrm{in}}$ as the absolute error of parameter estimate $P_{\mathrm{out}}$, we define light curve ``cases" in \autoref{tab:cases}.


\begin{table}
\caption{Cases with Modified Binary Classification}
    \begin{tabular}{cccc}
         \hline
         & Injected & Model & $\Delta P < 10\%$ \\
         & Sinusoid & Selection & \\
         \hline
         \hline
         Case 1 & $\checkmark$ & $\checkmark$ & $\checkmark$ \\
         
         Case 2 & $\checkmark$ & $\checkmark$ & \xmark \\
         
         Case 3 & $\checkmark$ & \xmark & $\checkmark$ \\
         
         Case 4 & $\checkmark$ & \xmark & \xmark \\
         
         Case 5 & \xmark & \xmark & ... \\
         
         Case 6 & \xmark & $\checkmark$ & ... \\
         \hline
         
    \end{tabular}

    \textbf{Note.} ``Model selection" column indicates whether or not model selection indicated the light curve contains evidence for periodicity.
    \label{tab:cases}
\end{table}

For the NBS, the model selection criterion is satisfied when the DRW+Sine model is preferred (typically, $\Delta\mathrm{BIC} < -2$). For Fourier methods, the criterion is satisfied when the power spectrum peak is significant (typically, $\left(1-\mathrm{FAP}\right) \geq 99.73\%$). At times, we also employ ``strict" model selection criterion ($\Delta\mathrm{BIC} < -6$ and $\left(1-\mathrm{FAP}\right) \geq 99.795\%$).

With $N_{i}$ as the number of case $i$ light curves, we define the following rates:

\begin{equation}\label{eq:tprp}
    \begin{split}
        \text{Genuine True Positive } \text{Rate (TPR($P$))} = & \\ \frac{N_{1}}{N_{1}+N_{2}+N_{3}+N_{4}},
    \end{split}
\end{equation}

with the ($P$) denoting the $P$ recovery condition, 

\begin{equation}\label{eq:fdr}
\begin{split}
    \text{Partial True Positive Rate (PPR) } = \\ \frac{N_{2}}{N_{1}+N_{2}+N_{3}+N_{4}},
\end{split}
\end{equation}

\begin{equation}\label{eq:tpr}
\begin{split}
    \text{True Positive Rate (TPR)} = \\ \frac{N_{1}+N_{2}}{N_{1}+N_{2}+N_{3}+N_{4}},
\end{split}
\end{equation}

\begin{equation}\label{eq:fpr}
    \text{False Positive Rate (FPR)} = \frac{N_{5}}{N_{5}+N_{6}}.
\end{equation}


Note, while one goal of this analysis is to constrain these rates, their exact values depend on the properties of our simulated data and our specific method of Bayesian and Fourier model selection. Thus, results found in this paper can inform SMBHB searches using the NBS, GLSP, and the WWZ, but exact rates cannot be extended to existing samples of SMBHB candidates, since these candidates were selected with completely different methods.
\section{Results} \label{sec:results}
We principally investigate the capacity of the NBS, GLSP, and WWZ methods to detect periodicity among AGN variability for a simulated population of SMBH and SMBHB light curves. Specifically, we identify the SMBHBs that can readily be detected in CRTS, ZTF, or LSST. Furthermore, given the expected volume of incoming LSST AGN light curves, we investigate the feasibility of a rapid triage using GLSP.

\subsection{Detection Capacity} \label{subsec:detection}

\begin{figure*}[ht!]
    \centering
    \includegraphics[width=1.\linewidth]{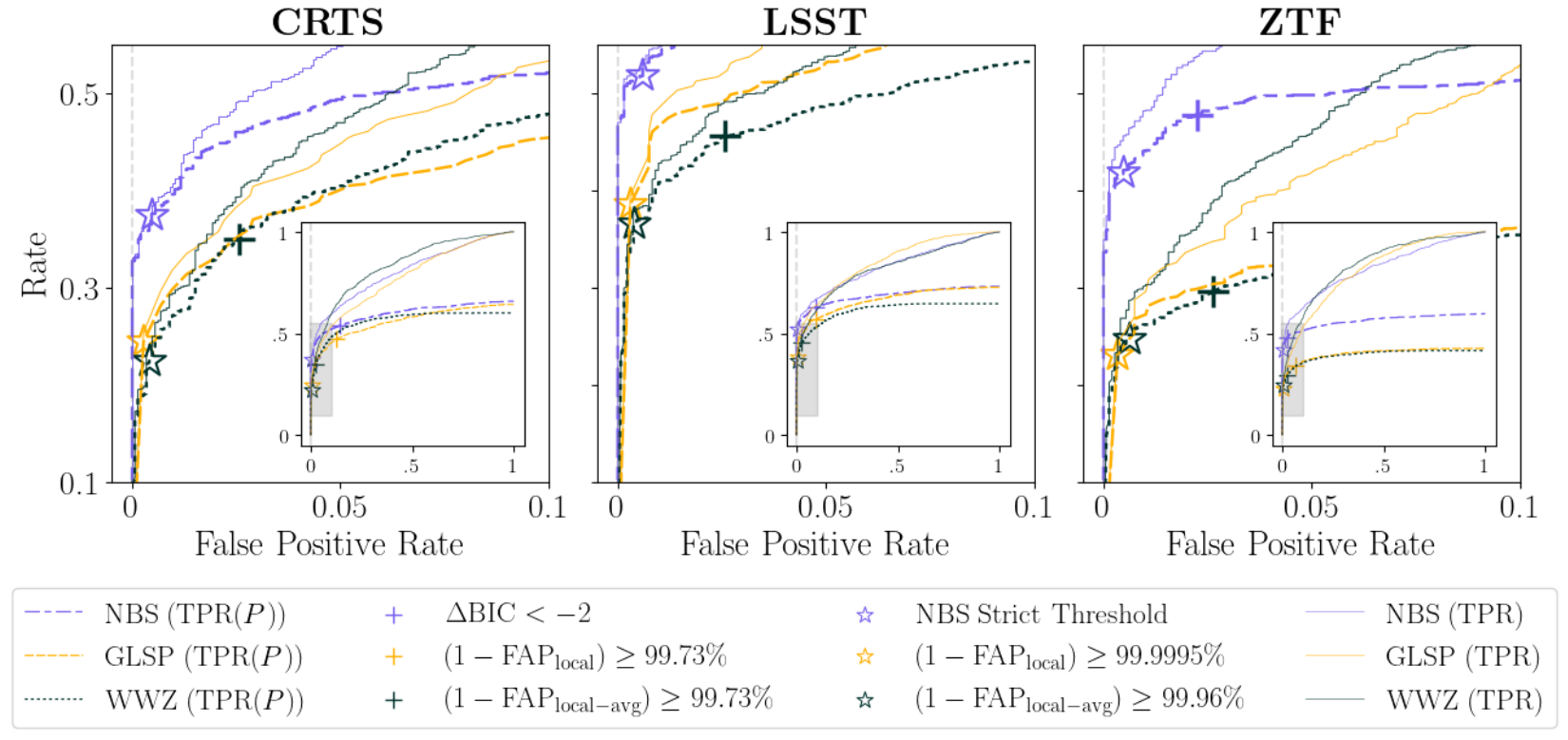}
    \caption{TPR-FPR (faint solid lines) and TPR($P$)-FPR curves (dashed/dotted lines) shown for each method's best classifier. Typical model selection thresholds (plus markers) produce complete samples of genuine true positives with relatively high false positive counts. Strict thresholds (open star markers) yield similar completeness with far fewer false positives. TPR($P$)s, FPRs and AUC values are shown in \autoref{tab:auc_tprp} and \autoref{tab:auc_tprp2}.}
    \label{fig:rate_plot_best}
\end{figure*}

To evaluate detection capacity, we observe the TPR($P$)s, TPRs and FPRs produced by each classifier. From here on, we describe samples with high TPR($P$)/TPRs as ``complete." Since these rates depend on which model selection threshold we use, we create recovery rate plots, which plot TPR or TPR($P$) versus\ FPR for different model selection thresholds. Since TPR versus\ FPR plots are receiver operating characteristic (ROC) curves, we quantitatively assess the performance of each classifier by computing the area under the TPR versus FPR curve, also known as the AUC value. In general, a larger AUC value indicates a better-performing classifier, as it equals the probability the classifier will rank a positive simulation better than a negative one \citep{fawcett_roc}. We use the TPR($P$) versus\ FPR curves to constrain TPR($P$)s and FPRs obtained using ``typical" literature-recommended thresholds ($\Delta\mathrm{BIC} < -2$ and $\left(1-\mathrm{FAP}\right)\geq99.73\%$). For certain classifiers, we also identify ``strict" model selection thresholds which produce very precise samples ($\mathrm{FPR} \lesssim .5\%$). Recorded values are displayed in \autoref{tab:auc_tprp} and \autoref{tab:auc_tprp2} and recovery rate plots for each method's best classifier are shown in \autoref{fig:rate_plot_best}.

\begin{table*}[ht!]
\centering
\caption{AUC values and Rates at Typical Model Selection Threshold for Each Classifier}
    \begin{tabular}{cccccccc}
        \hline
         {Method} & {Classifier} & & AUC values & & & (TPR($P$), FPR)\%\\
         & & CRTS & LSST & ZTF & CRTS & LSST & ZTF \\
         \hline
         \hline
         NBS & $\Delta$BIC & 0.808 & 0.836 & 0.825 & (53.96, 14.24) & (62.73, 9.25) & (47.68, 2.24) \\
         
         \hline
         & $\mathrm{FAP_{Gauss}}$ & 0.836 & 0.854 & 0.842 & ... & ... & ... \\
         GLSP & $\mathrm{FAP_{global}}$ & 0.856 & 0.880 & 0.870 & (17.90, 0.55) & (32.06, 0.68) & (19.10, 0.54) \\
         & $\mathrm{FAP_{local}}$ & 0.776 & 0.854 & 0.815 & (47.27, 12.79) & (57.15, 9.77) & (33.94, 6.30) \\
         
         \hline
         & $\mathrm{FAP_{global}}$ & 0.831 & 0.813 & 0.851 & (13.73, 0.48) & (15.99, 0.38) & (17.14, 0.54) \\
         WWZ & $\mathrm{FAP_{local\text{-}local}}$ & 0.774 & 0.807 & 0.789 & (45.15, 18.94) & (52.02, 18.57) & (32.37, 9.28) \\
         & $\mathrm{FAP_{local\text{-}avg}}$ & 0.837 & 0.823 & 0.841 & (34.97, 2.56) & (45.56, 2.56) & (29.63, 2.64) \\
         & $\mathrm{FAP_{avg\text{-}local}}$ & 0.791 & 0.785 & 0.777 & (11.34, 1.04) & (17.32, 1.05 & (12.95, 1.08) \\
         \hline
    \end{tabular}

    \vspace{.25cm}
    \textbf{Note.} Select TPR($P$)s and FPRs are marked by colored plus markers in \autoref{fig:rate_plot_best}. Most notably, the NBS $\Delta\mathrm{BIC}$, GLSP $\mathrm{FAP_{local}}$ and WWZ $\mathrm{FAP_{local\text{-}avg}}$ produce impressively high TPR($P$)s.
    \label{tab:auc_tprp}
\end{table*}

\begin{table*}[ht!]
\centering
\caption{TPR($P$)s and FPRs at Strict Model Selection Thresholds for Each Classifier}
    \begin{tabular}{cccccccc}
        \hline
         {Method} & {Classifier} & & Thresholds & & & (TPR($P$), FPR)\%\\
         & & CRTS & LSST & ZTF & CRTS & LSST & ZTF \\
         \hline
         \hline
         NBS & $\Delta$BIC & $-9$ & $-9$ & $-6$ & ($37.5$, $0.48$) & ($51.87$, $0.60$) & ($41.79$, $0.47$) \\
         
         \hline
         & $\mathrm{FAP_{Gauss}}$ & $10^{-25}$ & $10^{-60}$ & $10^{-110}$ & ($12.43$, $0.35$) & ($24.07$, $0.75$) & ($15.89$, $0.61$) \\
         GLSP & $\mathrm{FAP_{global}}$ & ... & ... & ... & ... & ... & ... \\
         & $\mathrm{FAP_{local}}$ & $5\times 10^{-5}$ & $5\times 10^{-5}$ & $5\times 10^{-5}$ & ($24.39$, $0.28$) & ($38.52$, $0.30$) & ($23.09$, $0.34$) \\
         
         \hline
         & $\mathrm{FAP_{global}}$ & ... & ... & ... & ... & ... & ... \\
         WWZ & $\mathrm{FAP_{local\text{-}local}}$ & $5\times 10^{-5}$ & $5\times 10^{-5}$ & $5\times 10^{-5}$ & ($20.49$, $0.76$) & ($35.66$, $1.20$) & ($22.04$, $0.61$) \\
         & $\mathrm{FAP_{local\text{-}avg}}$ & $4\times 10^{-4}$ & $4\times 10^{-4}$ & $4\times 10^{-4}$ & ($22.40$, $0.42$) & ($36.54$, $0.38$) & ($24.72$, $0.61$) \\
         & $\mathrm{FAP_{avg\text{-}local}}$ & ... & ... & ... & ... & ... & ... \\
         \hline
    \end{tabular}

    \vspace{.25cm}
    \textbf{Note.} Select TPR($P$)s and FPRs are marked by colored stars in \autoref{fig:rate_plot_best}. Using strict model selection thresholds for $\Delta\mathrm{BIC}$, $\mathrm{FAP_{local}}$ and $\mathrm{FAP_{local\text{-}avg}}$ significantly reduces FPRs with marginal impact to TPR($P$)s.
    \label{tab:auc_tprp2}
\end{table*}

\autoref{tab:auc_tprp2} and \autoref{fig:rate_plot_best} clearly illustrate the nested Bayesian sampler's $\Delta\mathrm{BIC}$ is the best classifier. Despite its relatively low AUC values, it produces incomparably complete samples with minimal false positives.

\autoref{tab:auc_tprp} and \autoref{tab:auc_tprp2} show that GLSP classifiers $\mathrm{FAP_{Gauss}}$, $\mathrm{FAP_{global}}$ and $\mathrm{FAP_{local}}$ are the next best classifiers. The $\mathrm{FAP_{Gauss}}$ distinguishes DRW and sinusoidal variability well despite assuming Gaussian noise, but still produces very incomplete samples. Still, thanks to its efficiency, we further investigate its use for a light curve triage in \autoref{subsec:triage}. The $\mathrm{FAP_{global}}$ classifier produces excellent AUC values and low FPRs, but is a clear second option. Using strict model selection thresholds, the $\mathrm{FAP_{local}}$ produces the most complete samples with the lowest FPRs. Overall, $\mathrm{FAP_{Gauss}}$ is best fit for a triage and the $\mathrm{FAP_{local}}$ classifier will provide the most complete samples of periodic quasars with minimal false positives.

The WWZ $\mathrm{FAP_{global}}$ scores high AUC values but, along with the $\mathrm{FAP_{avg\text{-}local}}$ classifier, it produces very incomplete, low-FPR samples. The $\mathrm{FAP_{local\text{-}local}}$ classifier produces relatively low AUC values and moderately complete (TPR($P$) $\approx$ 20--35$\%$) for low-FPR samples. Using a strict threshold, the $\mathrm{FAP_{local\text{-}avg}}$ produces the most complete samples with the fewest false positives. The $\mathrm{FAP_{local\text{-}avg}}$ is thus clearly the best WWZ classifier.

\begin{figure*}[ht!]
    \centering
    \includegraphics[width=1.\linewidth]{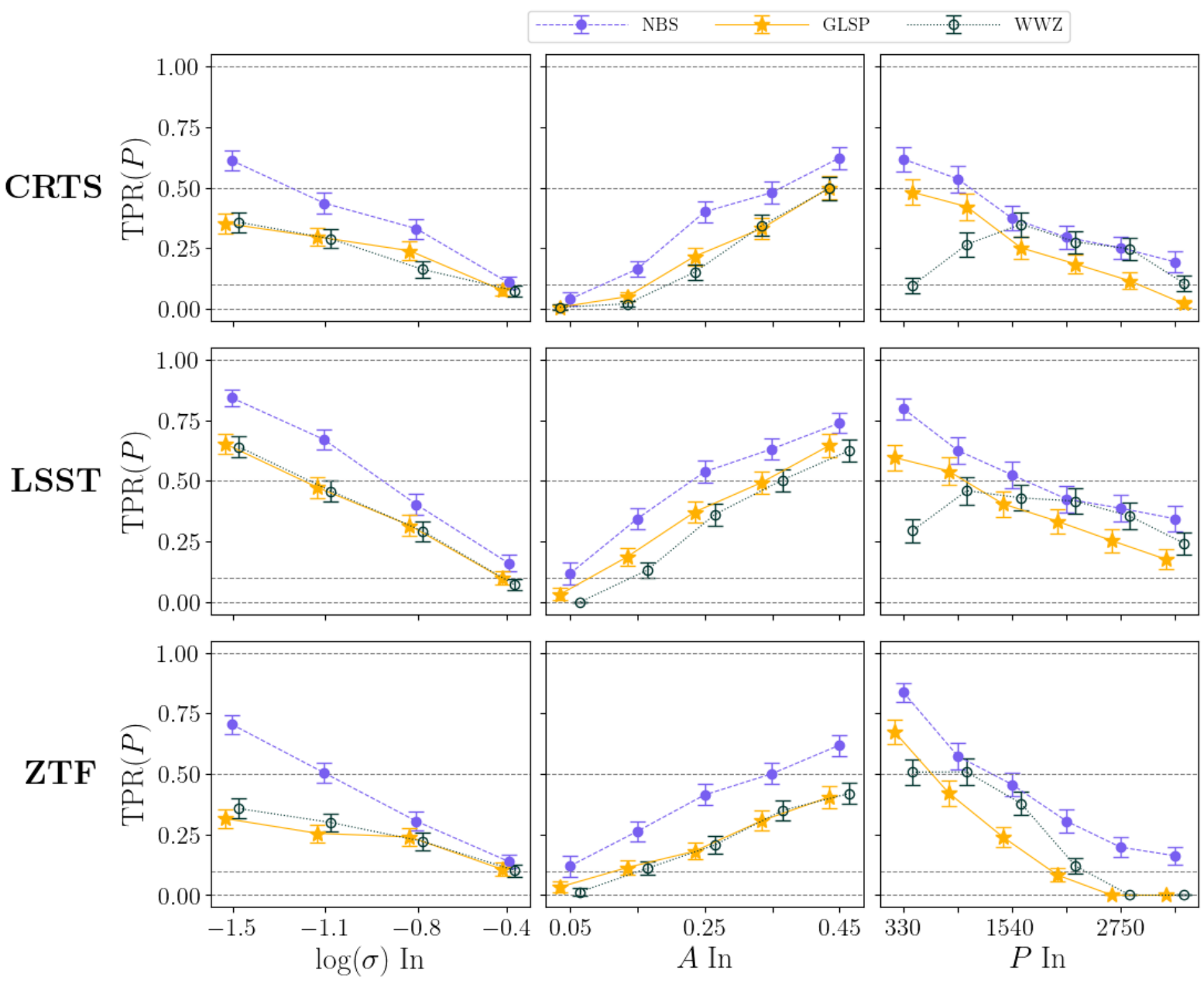}
    \caption{TPR($P$) for NBS (closed purple circles), GLSP (yellow stars), and WWZ (open green circles) plotted as functions of input signal parameters $\sigma_{\mathrm{in}}$, $A_{\mathrm{in}}$, and $P_{\mathrm{in}}$. TPR($P$)s are highest for LSST data (middle panel) and lowest for CRTS data (top panel). Offsets from bin centers (marked by x-axis tick marks) are used to show errorbars. Clearly, all methods favorably detect high-amplitude, short-period, low-DRW-variability light curves.}
    \label{fig:case_rates_all}
\end{figure*}

With the $\Delta\mathrm{BIC}$, $\mathrm{FAP_{local}}$ and $\mathrm{FAP_{local\text{-}avg}}$ as the representative classifiers for the NBS, GLSP and WWZ, respectively, we investigate which types of sinusoidal signals each method is most likely to detect. Here, we plot each methods' TPR($P$)s as functions of injected parameters $P_{\mathrm{in}}$, $A_{\mathrm{in}}$ and $\sigma_{\mathrm{in}}$ (\autoref{fig:case_rates_all}). We use TPR($P$)s acquired by imposing strict model selection thresholds (\autoref{tab:auc_tprp2}) on light curves which fall in each parameter bin. The error bars for each bin are calculated using the confidence interval for the binomial proportion

\begin{equation}
    \frac{n_{S}}{n} \pm \frac{z}{n \sqrt{n}}\sqrt{n_{S}n_{F}},
\end{equation}
where $n$ is the number of trials, $n_{S}$ is the number of successes, $n_{F}$ is the number of failures and $z$ is the 1-$\alpha$/2 quantile of a normal distribution \citep{binomial}.


\autoref{fig:case_rates_all} shows, for all three methods, our ability to detect periodicity is positively correlated with injected sinusoidal amplitude $A_{\mathrm{in}}$ and negatively correlated with injected DRW variability amplitude $\sigma_{\mathrm{in}}$. Specifically, only $\lesssim15\%$ of low-$A_{\mathrm{in}}$ light curves ($A_{\mathrm{in}} \leq 0.15$) are detected. The NBS detects a significant proportion ($\gtrsim50\%$) of light curves with high amplitude ($A \geq 0.25$) or low DRW variability ($\log\left(\sigma_{in}\right) < -1.0$). These correlations are identical for the GLSP $\mathrm{FAP_{local}}$ and WWZ $\mathrm{FAP_{local\text{-}avg}}$ classifiers, but they detect fewer light curves in every case. Though not shown, TPR($P$)s showed no significant correlation with the injected DRW timescale, $\tau_{\mathrm{in}}$.

\begin{figure*}
    \centering
    \includegraphics[width=.95\linewidth]{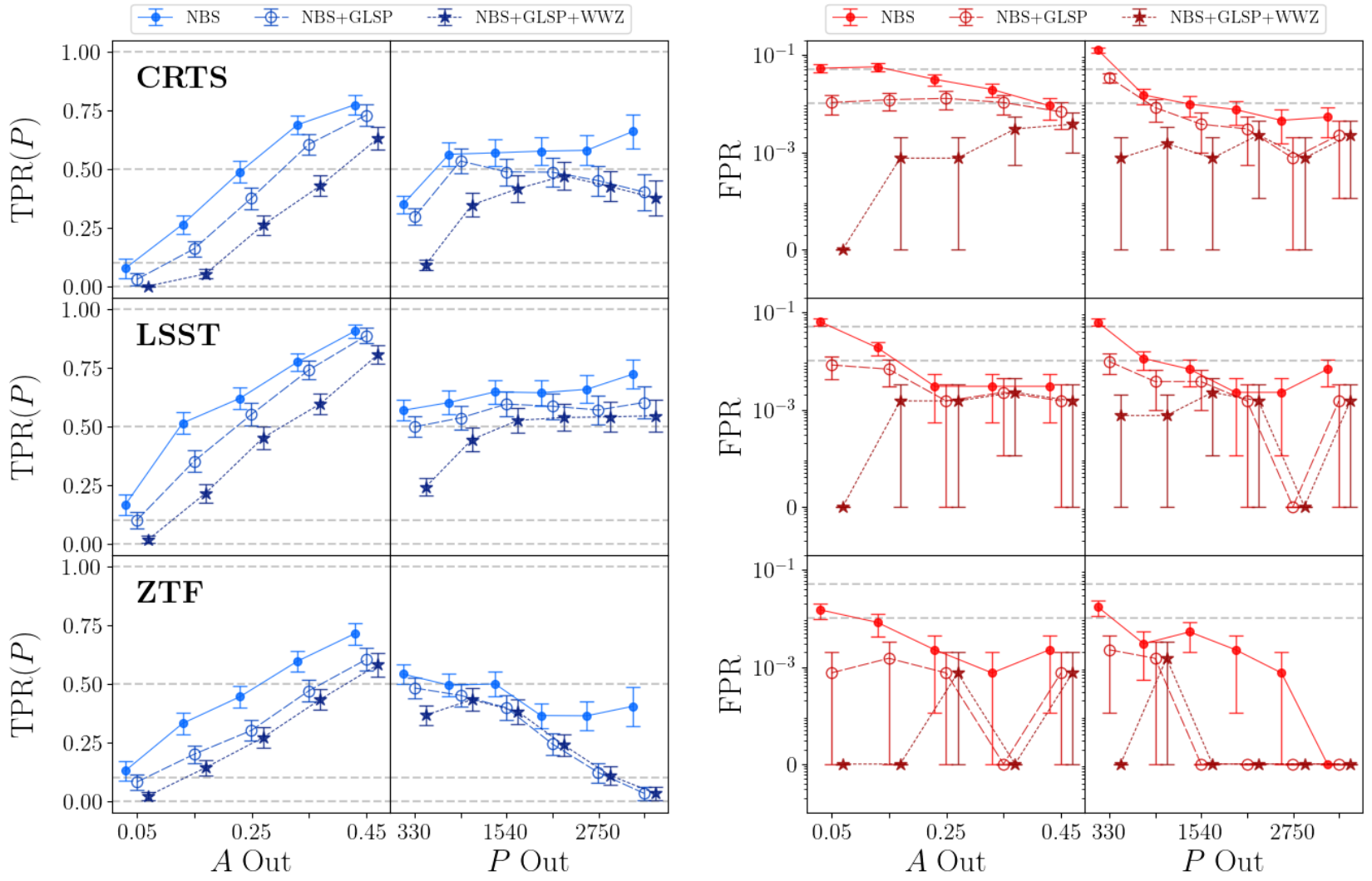}
    \caption{TPR($P$)s and (locally-normalized) FPRs for NBS (closed light-blue or red circles, solid lines), NBS+GLSP (open medium-blue or red circles, dashed lines), and NBS+GLSP+WWZ (dark-blue or red stars, dotted lines) model selection plotted against NBS outputs, $A_{\mathrm{out}}$ and $P_{\mathrm{out}}$. Offsets from bin centers (marked by x-axis tick marks) are used to show errorbars. Both NBS+GLSP and NBS+GLSP+WWZ joint model selection offer opportunities to scour samples of false positives while retaining a moderate TPR($P$).}
    \label{fig:tprp_fpr_aout_pout}
\end{figure*}

Each method's TPR($P$)s are also strongly correlated with the injected sinusoidal period. The NBS and GLSP TPR($P$)s decrease as the period approaches the observation baseline. Specifically, the NBS detects $\sim$35\% of LSST-like long-period ($P \geq 3000$) light curves and slightly fewer CRTS-like long-period light curves. This indicates light curves with $<2$ cycles of periodicity are detectable by the NBS, reinforcing findings from \citet{witt2022_agnvar_bayesian}. With evidence from \autoref{fig:tprp_fpr_aout_pout} which indicates that false positives tend to output low periods, this finding suggests a search condition like $P_{\mathrm{out}}/T_\mathrm{obs} \geq 1.5$ may exclude genuine true positives for little reduction to the FPR. The WWZ outperforms the GLSP for longer periods, but worsens for short-period ($P \lesssim 1000$) light curves. Likely, low-cadence intervals in light curve data produce spurious power spectrum peaks which disrupt parameter and detection significance estimates, most noticeably for short-timescale periodicity. Overall, we expect the GLSP and NBS to detect $\sim$50--80$\%$ of short-period light curves. This is especially promising for future application to very short-period ($\sim$1--3 days) light curves corresponding to LISA-detectable binary black holes.

To this point, we have presented our results with respect to input parameters in order to assess, all else equal, the relative detection capacity of each method for a variety of sinusoid signals. In real data applications, our primary concern will be filtering out non-periodic quasars. Here we investigate how combining each method's model selection (and additional output parameters) can clear samples of false positive non-periodic quasars while retaining a reasonable portion of periodic quasars. With $\mathbf{s}_{\mathrm{best}}$ as the best-fit signal vector from the NBS DRW+Sine model fit, we let

\begin{equation}
    \mathrm{(S/N)_{NBS}} = \mathbf{s}_{\mathrm{best}}^{T} \cdot C^{-1} \cdot \mathbf{s}_{\mathrm{best}}.
    \label{eq:snr_nbs}
\end{equation}

\begin{figure*}
    \centering
    \includegraphics[width=1.\linewidth]{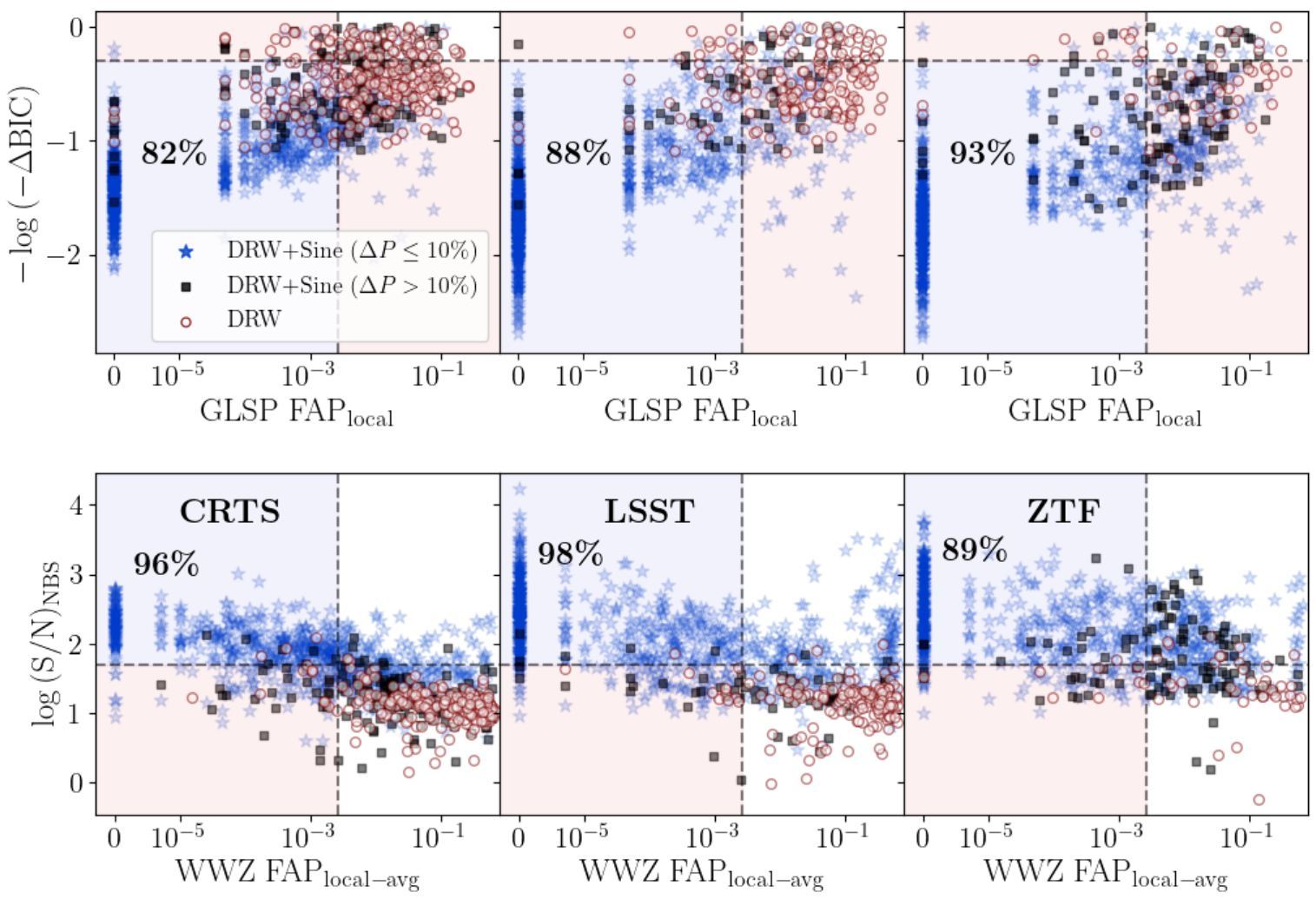}
    \caption{Log-scaled $\Delta\mathrm{BIC}$ vs. GLSP $\mathrm{FAP_{local}}$ (top) and $\mathrm{(S/N)_{NBS}}$ vs. WWZ $\mathrm{FAP_{local\text{-}avg}}$ (bottom) for genuine true positives (blue stars), partial true positives (black squares) and false positives (red open circles). Dashed lines mark $\Delta\mathrm{BIC} = -2$, $ (1-\mathrm{FAP_{local}}) = 99.73\%$, $\mathrm{FAP_{local\text{-}avg}} = 99.73\%$, $\mathrm{(S/N)_{NBS}} = 50$. Blue stars greatly outnumber red and green stars in the blue-shaded regions (see blue star occurrence rate in bold text), indicating that joint model selection preferentially removes false positives and partial true positives while retaining high TPR($P$).}
    \label{fig:gdi}
\end{figure*}

In \autoref{fig:gdi} we present model selection outputs $\Delta\mathrm{BIC}$, $\mathrm{FAP_{local}}$, $\mathrm{FAP_{local\text{-}avg}}$ and $\mathrm{(S/N)_{NBS}}$ (\autoref{eq:snr_nbs}) to visualize regions where genuine true positives, partial true positives, and false positives fall. Finally, in \autoref{fig:tprp_fpr_aout_pout} we plot the TPR($P$)s and locally-normalized FPRs\footnote{Locally-normalized $\mathrm{FPR}=\frac{\mathcal{N\left(\text{FPs in output bin}\right)}}{\mathcal{N\left(\text{DRW LCs in output bin}\right)}}$} acquired by combining NBS ($\Delta\mathrm{BIC}$), GLSP ($\mathrm{FAP_{local}}$), and WWZ ($\mathrm{FAP_{local\text{-}avg}}$) model selection against NBS outputs, $A_{\mathrm{out}}$ and $P_{\mathrm{out}}$. Using this, we note advantages and drawbacks of combining each method's model selection, and identify the ranges of expected $A_{out}$ and $P_{\mathrm{out}}$ which correspond to genuine true positives and false positives. For this plot, we use typical NBS, GLSP, and WWZ model selection thresholds. Though using strict thresholds are redundant for our sample size, we note that the following analysis may produce non-negligible FPRs for larger samples. In this case, strict thresholds may be ideal (see \autoref{subsec:class_roadmap} for a complete guide on how to use joint model selection for real data).

In the blue-shaded regions in \autoref{fig:gdi}, blue stars greatly outnumber red and green stars($\sim$95--99\% occurrence rate), indicating that joint model selection preferentially removes false positives and partial true positives while retaining high TPR($P$). We note that while the $\mathrm{(S/N)_{NBS}}>50$ criterion removes partial true positives and false positives from our samples, it also removes well-detected low-$A_{\mathrm{in}}/\sigma_{\mathrm{in}}$ true positives. We discuss the best and prudent uses of this and other aforementioned criteria in \autoref{subsec:class_roadmap}.

\autoref{fig:tprp_fpr_aout_pout} demonstrates that false positives tend to output low $A$ and low $P$. This further supports our earlier claim that the $T_{\mathrm{obs}}/P_{\mathrm{out}} \geq1.5$ detection criterion fails to filter out the problematic false positives. On the other hand, \autoref{fig:tprp_fpr_aout_pout} shows that combining GLSP and NBS model selection reduces these low-$A$ and low-$P$ false positive rates by more than an order of magnitude. Further incorporating WWZ model selection reduces the FPR even more, but at the expense of a much lower TPR($P$). Since we only run $\sim$3500 simulations, our FPR is only precise to $\sim$$10^{-3}$. As such, low ($\sim$0--10$^{-3}\%$) FPRs may correspond to high false positive counts for large samples. If this is the case, the minimal FPR returned with NBS+GLSP+WWZ model selection may justify the low TPR($P$).

\begin{figure}
    \centering
    \includegraphics[width=1\linewidth]{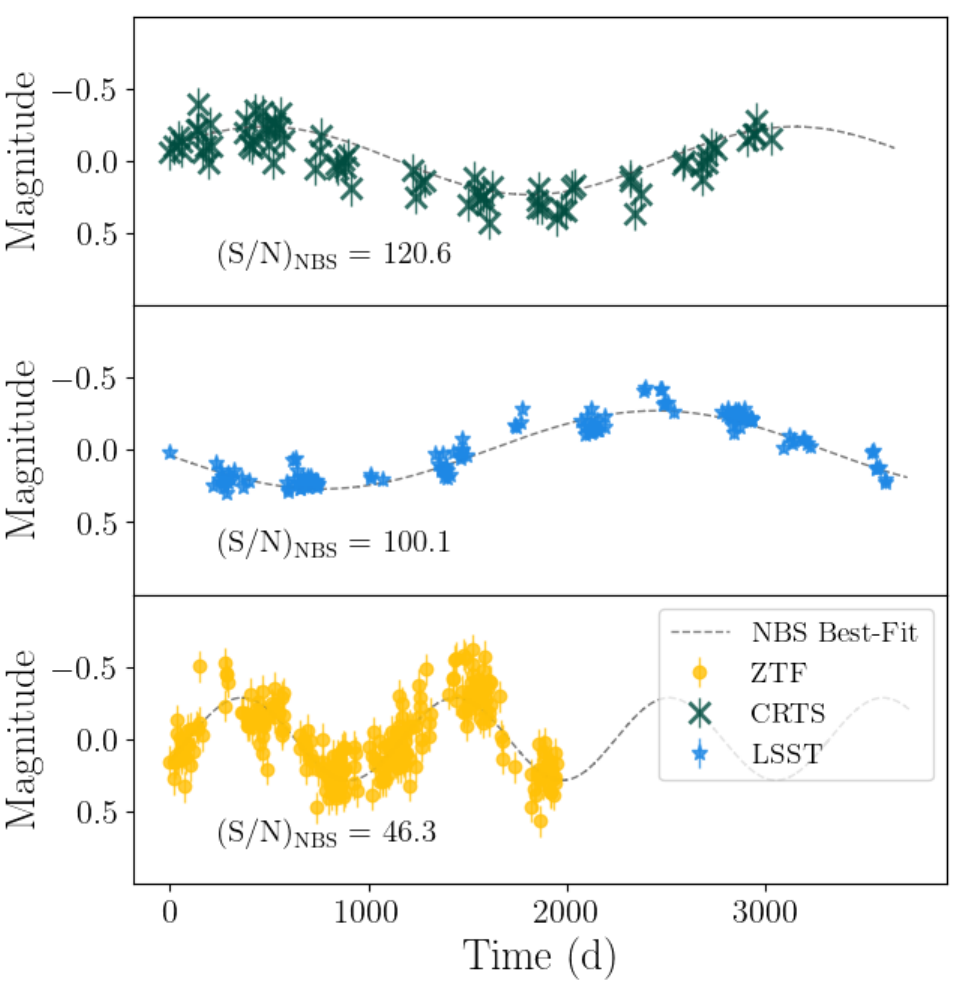}
    \caption{DRW light curves with $\Delta\mathrm{BIC} < -2$, $\mathrm{FAP_{local}} \geq 3\sigma$, and $\mathrm{FAP_{local\text{-}avg}} \geq 3\sigma$. 13, 9, and 4 DRW simulations satisfied all three criteria for CRTS data (pink X's), LSST data (blue stars) and ZTF data (beige circles) data, respectively. The simulations with the highest $\mathrm{(S/N)_{NBS}}$ are plotted here. NBS best fit sinusoid plotted in gray dashed lines. With at most ten years of data, we expect such convincing cases will inevitably pollute samples of periodic quasar candidates.}
    \label{fig:ex_bad_lc}
\end{figure}

We note that the very few DRW light curves which pass our combined model selection contain high-$\sigma$ DRW variations which mimic periodicity particularly well. There is no correlation with input DRW timescale $\tau$. Example light curves are shown in \autoref{fig:ex_bad_lc}. Given the stochastic nature of AGN variability, ruling out such cases will likely require data with longer baselines or validation via non-photometric methods.

\vspace{.35cm}
\subsection{LSST AGN Triage} \label{subsec:triage}

We expect LSST to produce $\geq1$ million quasar light curves with an extremely low ($\sim$$1/1000$) SMBHB candidate occurrence rate---the need for a computationally efficient triage is paramount. Given their minimal computational cost, we only consider the $\mathrm{FAP_{Gauss}}$ classifier and the $\mathrm{FAP}_{D_{\mathrm{local}}}$ classifiers as candidates for a periodic quasar triage. The latter is computed using the database methodology described in \autoref{subsec:rapid_fap} and is only applicable for light curves with extremely similar cadences and durations. Here, strict model selections are the highest model selection value which ensures FPR $\lesssim1\%$. Since there is no literature-recommended threshold for $\mathrm{FAP_{Gauss}}$, we only identify and use the strict model selection in this analysis.

\begin{figure}
    \centering
    \includegraphics[width=1.\linewidth]{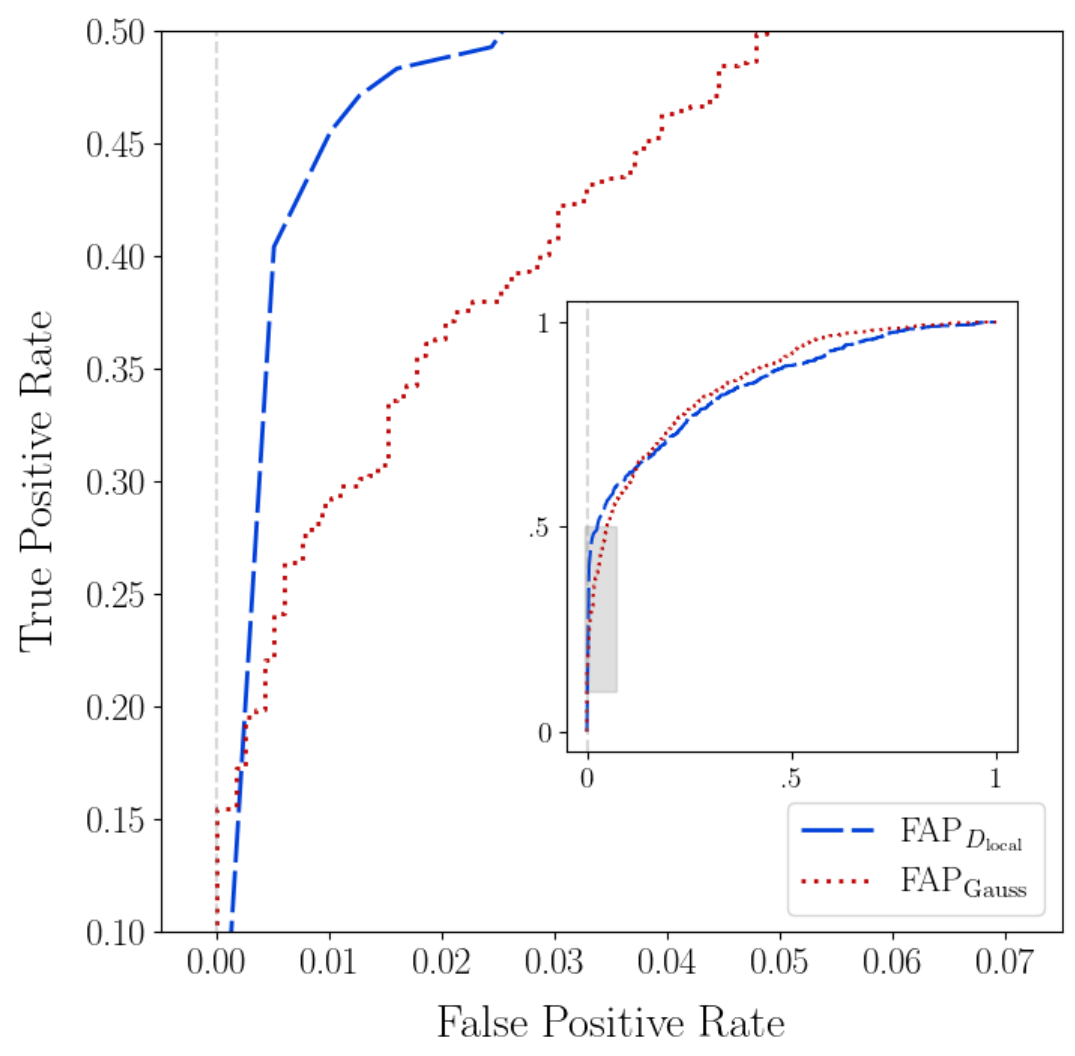}
    \caption{TPR-FPR plots shown for triage classifiers $\mathrm{FAP}_{\mathrm{Gauss}}$ (Red dotted line) and $\mathrm{FAP}_{D_{\mathrm{local}}}$ (Blue dashed line). Typical model selection thresholds are marked with pluses, and strict thresholds are marked with stars. Clearly, for a given FPR, the $\mathrm{FAP}_{D_{\mathrm{local}}}$ recovers more periodic light curves than the $\mathrm{FAP}_{\mathrm{Gauss}}$.}
    \label{fig:rate_plot_triage}
\end{figure}

\begin{table}[ht!]

\caption{AUC values and TPRs/FPRs at Typical and Strict Model Selection Thresholds for Triage Classifiers}
\centering
    \begin{tabular}{cccc}
        \hline
         {Classifier} & AUC & (TPR, FPR)\% & (TPR, FPR)\% \\
        & values & Typical & Strict \\
         \hline
         \hline
         $\mathrm{FAP_{Gauss}}$ & 0.859 & ... & (26.5, 0.8) \\
         
         $\mathrm{FAP}_{D_{\mathrm{local}}}$ & 0.844 & (60.0, 8.6) & (40.4, 0.5) \\
         
         \hline
    \end{tabular}

    \vspace{.25cm}
    \textbf{Note.} Recorded TPRs and FPRs, produced using typical (plus markers) and strict (star markers) model selection thresholds, are shown in \autoref{fig:rate_plot_triage}. Strict thresholds drawn from ``Thresholds $\to$ LSST" column of \autoref{tab:auc_tprp2}.
    \label{tab:auc_tpr}
\end{table}

\begin{figure*}
    \centering
    \includegraphics[width=1.\linewidth]{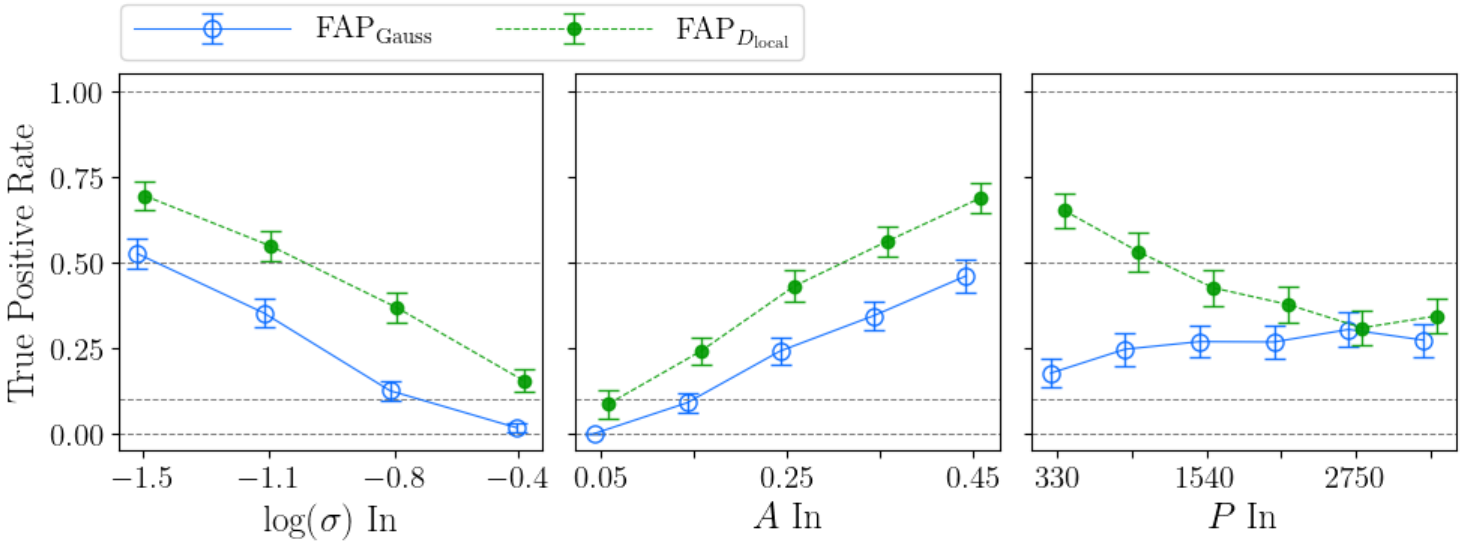}
    \caption{TPR for triage classifiers $\mathrm{FAP}_{\mathrm{Gauss}}$ (Blue x's, solid lines) and $\mathrm{FAP}_{D_{\mathrm{local}}}$ (Green circles, dotted lines) displayed as functions of input parameters $\log\left(\sigma\right)$, $A$, and $P$. Offsets from bin centers (marked by x-axis tick marks) are used to show errorbars. Triage classifiers favorably capture high-amplitude, low-DRW-variability light curves; $\mathrm{FAP}_{D_{\mathrm{local}}}$ also preferably captures short-period light curves.}
    \label{fig:case_rates_triage}
\end{figure*}

\begin{figure*}
    \centering
    \includegraphics[width=1.\linewidth]{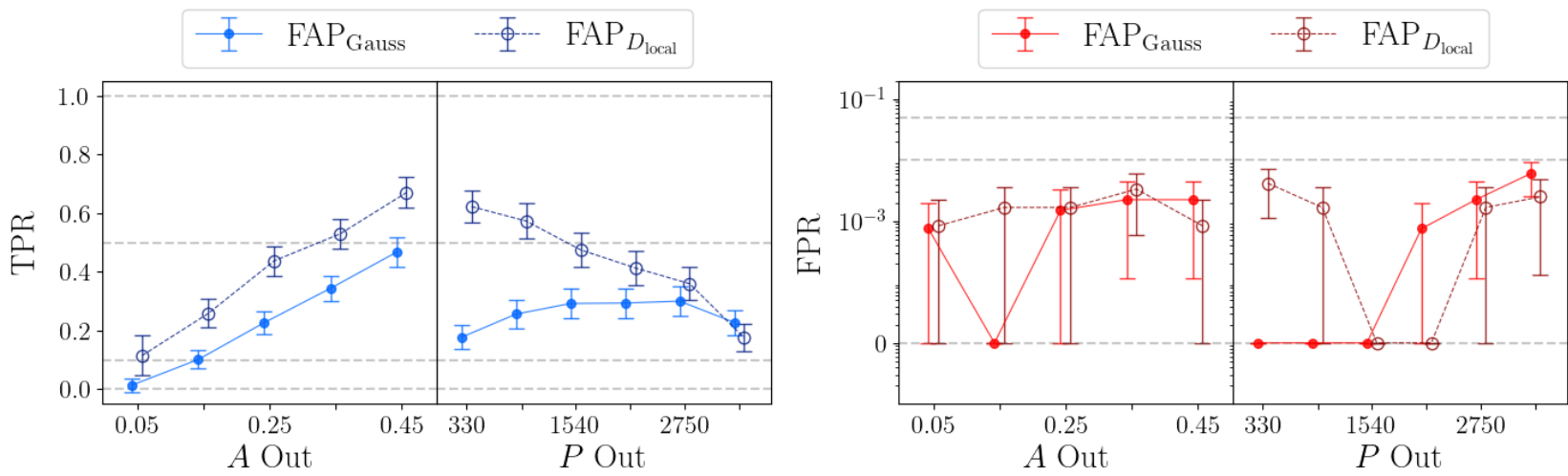}
    \caption{TPRs and (non-normalized) FPRs for triage classifiers $\mathrm{FAP}_{\mathrm{Gauss}}$ (Light blue circles, solid lines) and $\mathrm{FAP}_{D_{\mathrm{local}}}$ (Dark blue X's, dotted lines) plotted against GLSP outputs $A_{\mathrm{out}}$ and $P_{\mathrm{out}}$. Offsets from bin centers (marked by x-axis tick marks) are used to show errorbars. Here, $\mathrm{FAP}_{D_{\mathrm{local}}}$ produces a complete sample of periodic light curves with few false positives at high-amplitude and short-period outputs. $\mathrm{FAP}_{\mathrm{Gauss}}$ produces complete samples at high-amplitudes, but with a significant FPR for light curves with long-period outputs.}
    \label{fig:tpr_fpr_aout_pout_triage}
\end{figure*}

Impressively, the $\mathrm{FAP_{Gauss}}$ classifier scores a higher AUC value (\autoref{tab:auc_tpr}) than $\mathrm{FAP}_{D_{\mathrm{local}}}$. Despite this, \autoref{fig:rate_plot_triage} shows that for FPR $\lesssim5\%$ the $\mathrm{FAP}_{D_{\mathrm{local}}}$ classifier produces much more complete samples. In \autoref{fig:case_rates_triage}, nearly all TPRs follow the expected correlations established in \autoref{fig:case_rates_all}. The only exception is the lack of correlation between the $\mathrm{FAP_{Gauss}}$ classifier's TPR and injected period $P$. Where $\mathrm{FAP}_{D_{\mathrm{local}}}$ requires long-period peaks to exceed an uneven red noise background, the $\mathrm{FAP_{Gauss}}$ assumes an even Gaussian background and thus captures more long-period light curves. In \autoref{fig:tpr_fpr_aout_pout_triage}, we see that this failure to account for the red noise background causes $\mathrm{FAP_{Gauss}}$ false positives to output longer periods. For this reason, long-period candidates identified using a $\mathrm{FAP_{Gauss}}$ triage should be treated with caution. In general, we can expect the $\mathrm{FAP}_{D_{\mathrm{local}}}$ classifier to efficiently triage out false positives while retaining a moderate fraction of true positive candidates for follow-up observation. In the context of systematic LSST observations, the database computation the False Alarm Probability indicates a promising avenue for rapid handling of LSST quasar light curves.

\vspace{.25cm}
\section{Discussion} \label{sec:discussion}
\subsection{Previous and Ongoing Efforts}

In this paper, we investigated our ability to identify signatures of SMBHBs in AGN light curves in time-domain data sets, and compared the capabilities of a Bayesian classifier to frequency-domain identification techniques. In this process, we also compare the direct detection capabilities (measured via the TPR) with an additional hurdle of accurate identification of the binary period, which we quantify as the TPR(P), to expand over the more limited criteria in previous studies.

Numerous searches for SMBHBs have been carried out in existing time domain surveys \citep[][]{Graham2015_crts_candidate, Charisi2016_agnvar, liu2019_panstarrs_candidate, dgany2023_smbhb_ztf}, and other studies have placed predictions for LSST \citep[eg.][]{davis2024_agnvar_rubin}. However, few if any of these works have conducted a search with the triage method we recommend. With this tiered search methodology (first culling null results with fast fourier analyses, followed by more accurate NBS analyses), we expect to open significant possibilities for SMBHB searches in future data-heavy surveys. We will apply these methods to existing data sets, as well as more detailed simulated data, in future work.


\subsection{Model Development}\label{subsec:model_improvement}


Additional sources of periodically variable brightness in SMBHBs include: (1) Periodically variable accretion rates due to CBD-mini-disk interactions, (2) Periodic peaks for aligned self-lensing, and (3) sawtooth periodicity for eccentric binaries. Future work will simulate these additional signals with the \texttt{binlite} package \citet{dorazio2024_binlite} and develop corresponding likelihoods for model-fitting. Incorporating these additional signals may open the door to identifying more SMBHB candidates, but should be introduced with careful treatment of degeneracies between periodic signals at different harmonics of the orbital frequency.

Periodic signals from RDB and self-lensing in SMBHBs will have the same period across multiple bands. As such, correlating observations in multiple bands would multiply our effective cadence and could bolster parameter estimation. However, spectroscopic AGN observations indicate quasars change brightness \textit{and} color, implying light curves in different bands are subject to different, but correlated noise processes. In future work, we will incorporate models of multiband AGN ``time-lags" and quasar color changes to correlate AGN light curves in different bands and identify their common periodic signals.

Lastly, in the bootstrap calculation of the $\mathrm{FAP_{local}}$, we simulate DRW light curves using log-uniform distributions of $\sigma$ and $\tau$, potentially misrepresenting the nature of the candidate's stochastic variability. By using NBS posteriors of $\sigma$ and $\tau$ to simulate bootstrap light curves, we can more accurately assess detection significance, particularly in the event of low-$\sigma$ stochastic variability. As this is computationally expensive, we consider incorporating this methodology in follow-up analysis of exceptional candidates.



\subsection{Multi-messenger Astrophysics Opportunities}



As a key method for the detection of SMBHBs with EM observations, identification of binary-induced periodic behavior in AGN light curves provides an unparalleled opportunity for multi-messenger astrophysics. SMBHBs in the local universe are expected to emit low-frequency GWs detectable with pulsar timing arrays \citep[PTAs;][]{Burke-Spolaor2019}. The months-to-years long orbital periods probed by sinusoidal variability searches align perfectly with the most sensitive region of the PTA sensitivity band \citep[][]{xin2021_mma_crts}. 
With evidence of a stochastic GW background now present in PTA data sets \citep{nano15gwb, ipta_gwb}, GWs from individual SMBHBs may soon be detected by GW observatories.

For EM-emitting binaries, these developments have a few key implications. Firstly, if SMBHBs are indeed the source of the GW background, measurements of the amplitude of the background will directly inform upon the number density of SMBHBs within the local universe and the number of SMBHBs that surveys like LSST will uncover \citep{ng15smbbh, witt_gwblsst_prep}. As SMBHB candidates continue to be discovered within time-domain EM surveys, these numbers will provide a direct comparison of the local SMBHB number density from independent PTA methods.

Secondly, as EM candidates are uncovered with higher confidence, they can be searched for directly in PTA data sets as a multi-messenger analysis \citep{3c66b}. Even without a CW detection, multi-messenger techniques can place tighter constraints on the chirp-mass of individual SMBHBs \citep{3c66b}, and once a detection is made, will constrain the source parameters faster \citep{charisi2022_agnvar} and more accurately \citep{liu2021_mma}. \citet{charisi2025_mma} shows that using a joint EM-GW likelihood to simultaneously fit light curve and PTA data in a unified multi-messenger model improves parameter estimation even more. Future work will implement methods from this work into this proof-of-concept and further investigate how to improve their PTA analysis to construct a multi-messenger SMBHB search pipeline.

\subsection{Real Data Application}\label{subsec:class_roadmap}

\begin{figure}
    \centering
    \includegraphics[width=1.0\linewidth]{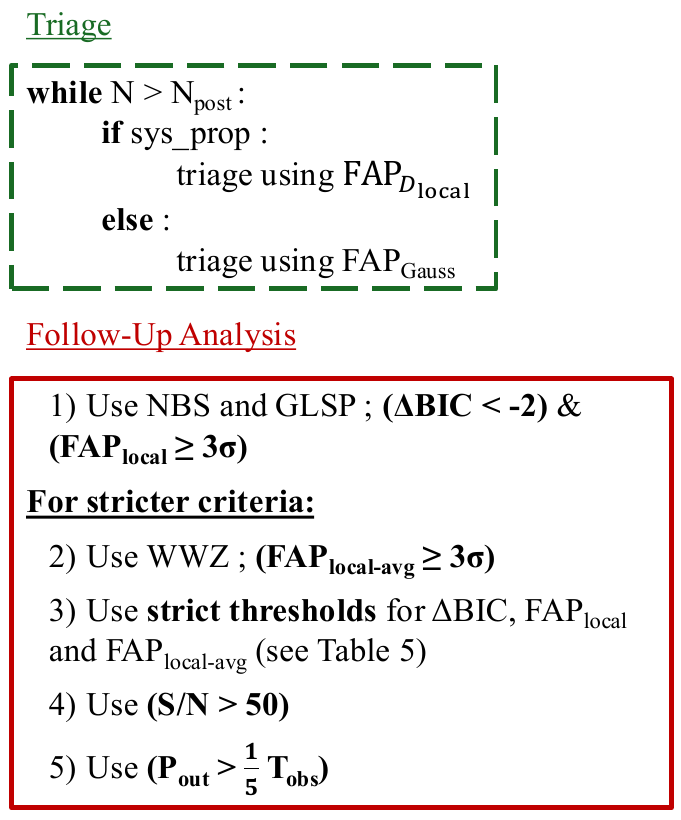}
    \caption{Data processing road map. $\mathrm{N}$ is the light curve count in the initial sample and $\mathrm{N_{post}}$ is the desired light curve count in the post-triage sample. ``$\mathrm{sys}$\textunderscore$\mathrm{prop}$" boolean should only be True for light curve samples with extremely similar cadencing (\autoref{subsubsec:surveys}). For triage, use thresholds given in \autoref{tab:auc_tprp} and \autoref{tab:auc_tprp2}. For follow-up analysis, ``stricter criteria" are provided in order of ascending strictness, so those light curves which pass all criteria should be given highest priority.}
    \label{fig:roadmap}
\end{figure}

Here, we present a guide for a real-dataset SMBHB search (\autoref{fig:roadmap}). Users can calculate their desired post-triage sample size $\mathrm{N_{post}}$, keeping in mind that a full NBS+GLSP follow-up analysis requires $\sim$2 hours of CPU time per light curve. Based on our analysis in \autoref{sec:results}, we expect that each criterion in the ``For stricter criteria" list preferentially removes non-periodic light curves from a given dataset. Follow-up priority should thus be assigned to a given light curve in proportion to the number of criteria it satisfies. If following the follow-up guide trims the dataset below the desired volume, relax the listed criteria from the bottom up.


\section{Conclusions} \label{sec:conclusions}
In this paper, we investigated our ability to photometrically detect SMBHBs in optical AGN light curves. We simulated a population of CRTS-, LSST-, and ZTF-like SMBH and SMBHB observations using a DRW model (and additional sinusoidal variability for the binaries). We analyzed each light curve with the NBS, GLSP, and WWZ in order to determine their capacity for detecting a periodic signal within stochastic variability. For the former, we performed Bayesian model selection using the $\Delta\mathrm{BIC}$ metric. For the GLSP and WWZ, we performed Fourier model selection using various calculations of the False Alarm Probability. Our most significant findings are:

\begin{itemize}
    \item Individually, the NBS most accurately recovers sinusoidal parameters and classifies DRW and DRW+Sine light curves.
    \item Combining GLSP model selection with NBS model selection retains $\sim$85--90$\%$ of NBS genuine true positives and only $\sim$20--30$\%$ of false positives.
    \item An abbreviated calculation of the FAP (\autoref{subsec:rapid_fap}) enables us to efficiently triage (TPR $\approx40\%$ and FPR $\approx.5\%$) LSST quasar light curves with minimal computation time ($\sim$$10^{7}$ light curves in $\sim$2--3\,days of CPU time).
\end{itemize}

Based on the previous three points, our primary recommendation for SMBHB searches in real data is to process large datasets ($>>$10$^{4}$ light curves) with a GLSP triage, then apply joint NBS+GLSP model selection ($\Delta\mathrm{BIC}$ and $\mathrm{FAP_{local}}$). The type of GLSP triage ($\mathrm{FAP}_{D_{\mathrm{local}}}$ or $\mathrm{FAP_{Gauss}}$) depends on the light curves' observational properties. See \autoref{subsec:class_roadmap} for a complete guide.

\begin{itemize}
    \item NBS and GLSP best detect short-period ($\sim$30--1000 d) periodic quasars, while WWZ best detects mid-to-long period ($\sim$1500--3500 d) light curves. All three methods are also most likely to detect high-amplitude, low-intrinsic-variability light curves. FPRs with all three methods are positively correlated with injected DRW variability.
    \item We predict that simply using $\sim$4--8 representative light curves in the database procedure will account for error introduced by differences in the photometric uncertainty of representative and candidate light curves.
\end{itemize}

For our methods, we find that survey baseline and photometric uncertainty are particularly important for detecting sinusoidal variability and filtering out false positives. LSST, beginning this year, will produce a 10-yr duration dataset with unprecedented photometric precision, making it incredibly amenable to SMBHB recovery based on our results. At the brink of this revolutionary data release, our recommendations instruct on best practices to identify these rare but incredibly important scientific systems in LSST data.

\acknowledgements
This material is based upon work supported by the National Science Foundation (NSF) under Grant No. AST-2149425, a Research Experiences for Undergraduates (REU) grant awarded to CIERA at Northwestern University. This research was supported in part through the computational resources and staff contributions provided for the Quest high performance computing facility at Northwestern University which is jointly supported by the Office of the Provost, the Office for Research, and Northwestern University Information Technology. CAW acknowledges support from CIERA, the Adler Planetarium, and the Brinson Foundation through a CIERA-Adler postdoctoral fellowship. 

AAM is partially supported by DoE award \#\,DE-SC0025599 and Cottrell Scholar Award \#\,CS-CSA-2025-059 from Research Corporation for Science Advancement.

\bibliography{agn}

\appendix
\section{Use and Accuracy of $\mathrm{FAP}_{D}$ for LSST light curves}\label{sec:appendix}

The calculation of $\mathrm{FAP}_{B}$ builds distributions of periodogram values using red noise simulations whose observational properties---photometric uncertainty, sampling cadence, and baseline---are taken from the candidate light curve. The calculation of $\mathrm{FAP}_{D}$ follows this procedure, but with red noise simulations whose observational properties are \textit{instead} drawn from one of few ``representative" light curves. Following this procedure inevitably alters the periodogram value distributions and subsequent FAP calculation. Here, we illustrate LSST survey windows of AGN will be sufficiently similar (\autoref{fig:db_pvs} top panel) such that using two representative LSST light curves ensures the database FAP calculation consistently reproduces the bootstrap FAP value. Additionally, we show large differences in $\mathrm{FAP}_{B}$ and $\mathrm{FAP}_{D}$ can nearly always be attributed to differences in the photometric uncertainty of the candidate light curve and representative light curve.

\begin{figure}[ht!]
    \centering
    \includegraphics[width=0.5\linewidth]{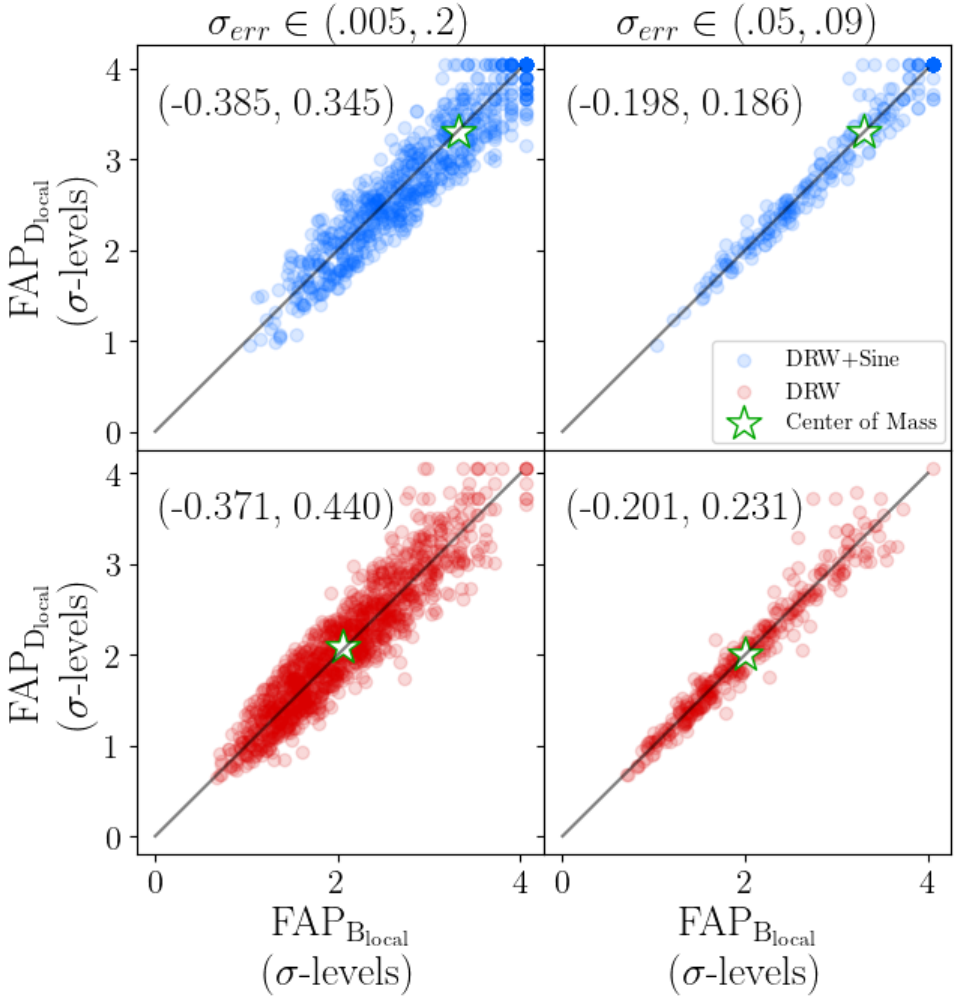}
    \caption{GLSP $\mathrm{FAP_{local}}$ calculated using the bootstrap procedure (x-axis) and database procedure (y-axis). FAP values are displayed in terms of the corresponding $\sigma$-level detections. 2-dimensional means, or ``center of masses" are marked by the white stars with green outline. The right two panels show these values for light curves with $\sigma_{err} \approx .07$, the $\sigma_{err}$ value for simulations in the database procedure. The top two panels plot these values for DRW+Sine light curves (blue circles), and the bottom two panels plot them for DRW-only light curves (red circles). Clearly, $\mathrm{FAP}_{D_{\mathrm{local}}}$ systematically reproduces $\mathrm{FAP}_{B_{\mathrm{local}}}$ to high accuracy, most noticeably for light curves with photometric uncertainty equal to that of the representative light curves.}
    \label{fig:fap_db_comp}
\end{figure}

\begin{figure}[ht!]
    \centering
    \includegraphics[width=1.\linewidth]{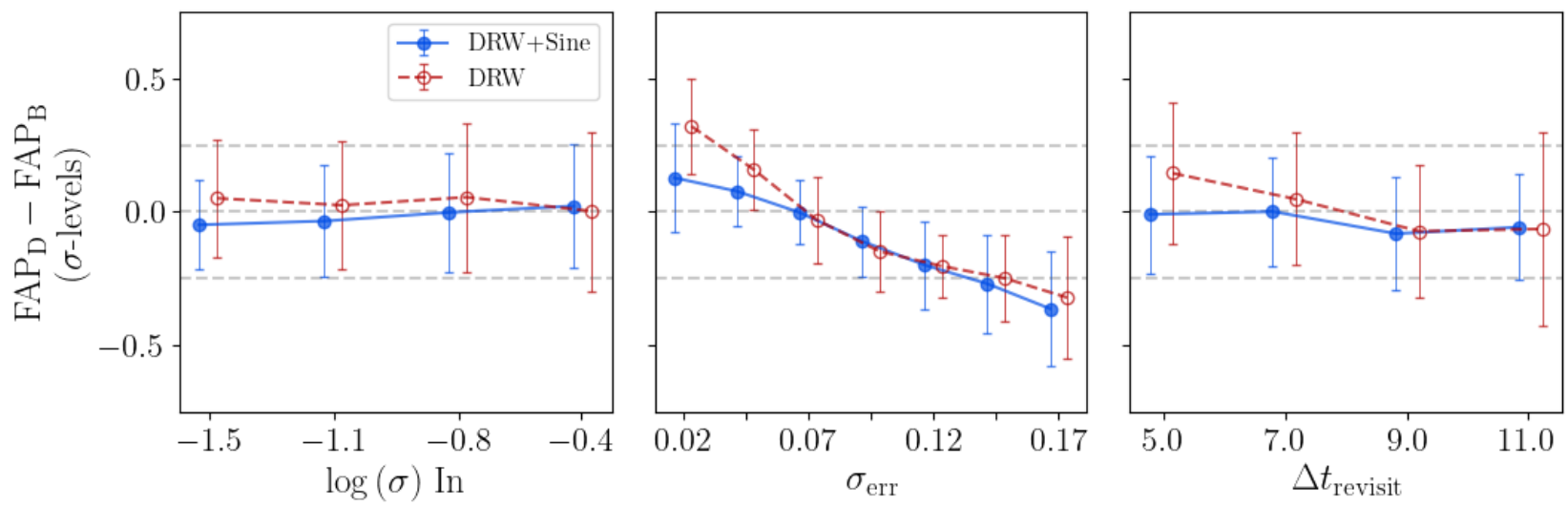}
    \caption{$\mathrm{FAP}_{D_\mathrm{local}}-\mathrm{FAP}_{B_{\mathrm{local}}}$ for DRW (open red circles) and DRW+Sine (closed blue circles) light curves vs. input parameters $\log\left(\sigma\right)$, $\sigma_{err}$ and $\Delta t_{\mathrm{revisit}}$. Errorbars are one standard deviation intervals about the mean. It is clear the database procedure is least accurate for light curves whose photometric uncertainty largely differ from the photometric uncertainty of the representative light curve.}
    \label{fig:fap_db_comp2}
\end{figure}

We draw representative LSST light curves from each of the numbered ``groups" in the top panel of \autoref{fig:db_pvs}, with their red noise periodogram distributions shown in the bottom panel. We use photometric uncertainty $\sigma_{err_{LSST}} \approx 0.07$ mag (the average LSST photometric uncertainty in our simulations) for each representative light curve. Here, we plot $\mathrm{FAP}_{B_{\mathrm{local}}}$ vs.\  $\mathrm{FAP}_{D_{\mathrm{local}}}$ to visualize their tight overall correlation. We also plot $\mathrm{FAP}_{D_\mathrm{local}}-\mathrm{FAP}_{B_{\mathrm{local}}}$ against five input parameters to identify which parameters are responsible for deviations between $\mathrm{FAP}_{D_{\mathrm{local}}}$ and  $\mathrm{FAP}_{B_{\mathrm{local}}}$. As seen in \autoref{fig:fap_db_comp}, the spread between $\mathrm{FAP}_{B_{\mathrm{local}}}$ and $\mathrm{FAP}_{D_{\mathrm{local}}}$ is minimal. The right panels of this figure show, for light curves with $\sigma_{err} \approx\sigma_{err_{LSST}}$, $\mathrm{FAP}_{B_{\mathrm{local}}}$ and $\mathrm{FAP}_{D_{\mathrm{local}}}$ are even more tightly correlated. Furthermore, \autoref{fig:fap_db_comp2} shows deviations between $\mathrm{FAP}_{B_{\mathrm{local}}}$ and $\mathrm{FAP}_{D_{\mathrm{local}}}$ most frequently occur when $\left|\sigma_{err}-\sigma_{err_{LSST}}\right|$ is greatest. We conclude the more closely our representative light curve's photometric uncertainty matches the candidate light curve's photometric uncertainty, the more faithfully $\mathrm{FAP}_{D_{\mathrm{local}}}$ will reproduce $\mathrm{FAP}_{B_{\mathrm{local}}}$. This suggests further breaking up our two ``groups" (see \autoref{fig:db_pvs}) into groups with low vs. high photometric uncertainty (creating four representative light curves instead of just two), we could ensure $\mathrm{FAP}_{D_{\mathrm{local}}}$ even \textit{more} closely matches $\mathrm{FAP}_{B_{\mathrm{local}}}$.

We argue, instead of exactly using the candidate light curve's survey window to compute the FAP (i.e., the bootstrap procedure), we can use survey windows from representative light curves to massively reduce computation time. Since LSST observational properties are sufficiently similar, using $\sim2-4$ representative light curves would suffice to accurately approximate $\mathrm{FAP}_{B}$ and estimate detection significance. Whereas calculating $\mathrm{FAP_{B}}$ for $10^{7}$ candidate light curves would take $\sim$2.5--5.5$\times10^{7}$\,hr of CPU time, the database procedure would require $\sim$12--24\,hr of CPU time for a small sacrifice to the accuracy of our detection significance estimate.

\end{document}